\documentclass[traditabstract]{aa}
\usepackage{graphicx}
\usepackage{txfonts}
\usepackage{natbib}
\bibpunct{(}{)}{;}{a}{}{,}
\usepackage{longtable}

\begin{document}

\title{A slitless spectroscopic survey for quasars near quasars
  \thanks{Based on observations collected at the European Southern Observatory,
          Chile (Proposals 70.A-0425, 074.A-0273 and 075.A-0141). Data taken
          under proposals 68.A-0330 and 70.A-0384 were obtained from the ESO
          Science Archive.}
}
\titlerunning{A slitless spectroscopic survey for quasars near quasars}
\author{G.~Worseck \inst{1} 
  \and L.~Wisotzki \inst{1} 
  \and F.~Selman \inst{2}}
\institute{Astrophysikalisches Institut Potsdam, An der Sternwarte 16, 14482 Potsdam, Germany\\
           \email{gworseck@aip.de}
	   \and
           European Southern Observatory, Alonso de Cordova 3107, Casilla 19001, Vitacura, Santiago, Chile}
\date{Received 9 May 2008 / Accepted 10 June 2008}

\abstract{
We present the results of the ``Quasars near Quasars'' (QNQ) survey, a CCD-based
slitless spectroscopic survey for faint $V\la 22$ quasars at $1.7\la z\la 3.6$
on 18 $26\farcm 2\times 33\farcm 5$ fields centred on bright quasars at
$2.76<z<4.69$. In total 169 quasar candidates with emission lines were selected
from the extracted flux-calibrated spectra on the basis of well-defined automatic
selection criteria followed by visual inspection and verification. With follow-up
spectroscopy of 81 candidates that were likely to reside at $z>1.7$ we were able
to confirm 80 new quasars at $0.580\le z\le 3.586$ on 16 of our fields. 64 of the
newly discovered quasars are located at $z>1.7$. The overall high success rate
implies that most of the remaining 88 candidates are quasars as well, although
the majority of them likely resides at $z<1.7$ on the basis of the observed line
shapes and strengths. Due to the insufficient depth of the input source
catalogues needed for extraction of the slitless spectra our survey is not well
defined in terms of limiting magnitude for faint $2.5\la z\la 3.6$ quasars whose
Ly$\alpha$ emission is detectable well beyond $V=22$, albeit at a continuum
$S/N\la 1$. While not useful for characterising the evolving space density of
quasars, our sample provides many new closely spaced quasar sightlines around
intensely studied quasars for further investigations on the three-dimensional
distribution of the intergalactic medium.

}
\maketitle

\section{Introduction}

Recent large optical surveys, such as the Sloan Digital Sky Survey (SDSS) and
the 2dF QSO Redshift Survey (2QZ), have revealed thousands of previously
unknown quasars selected on the basis of their broadband optical colours
\citep{schneider07,croom04}.  Colour selection is efficient if quasar
candidates are well separated from normal stars in multidimensional colour
space, most notably at $z\la 2.2$ (UV excess) and at $z\ga 3.5$. However, even
optical multicolour surveys are systematically incomplete at $2.5\la z\la
3.5$, where the colours of quasars and stars are similar
\citep[e.g.][]{warren91a,richards02b}. Incompleteness in this redshift range
can be significantly reduced by a better tracing of the spectral energy
distributions with additional filters, e.g.\ by incorporating mediumband
filters as in the COMBO-17 survey \citep{wolf03}.

Alternatively, slitless spectroscopy is a particularly efficient way to
find quasars at redshifts $z\ga 2$ because of the prominent Ly$\alpha$ emission
line redshifted into the optical wavelength regime. Early surveys 
generated hundreds of quasars by visual scanning of objective-prism
photographic plates for emission-line objects
\citep[e.g.][]{osmer80b,crampton85}.
Subjecting such plates to digitisation with fast measuring machines 
made it possible to employ the automated selection of quasar candidates
\citep{clowes84,hewett85} and to build substantial quasar samples 
at redshifts $0 \la z \la 3.2$ \citep{hewett95,wisotzki00}.
The systematic CCD-based slitless survey for $2.7\la z\la 4.8$ quasars
by \citet{schneider94} was among the first to quantify the declining space 
density of high-redshift quasars \citep{schmidt95}. 

Apparent pairs or close groups of high-redshift quasars are very attractive
targets to study the three-dimensional distribution of the intergalactic medium
(IGM). But high-resolution studies have so far been limited to a small number
of suitable groups of bright quasars \citep[e.g.][]{dodorico02,dodorico06}.
Going fainter than $V\sim 19$ immediately limits the achievable spectral
resolution and S/N. A possible compromise lies in combining high-resolution
spectra of bright quasars with lower resolution, lower S/N data of fainter
quasars in the surroundings. \citet{pichon01} argued that one can significantly
improve the recovery of the 3-dimensional topology of the IGM this way.

Although most known quasars were colour-selected from either SDSS or 2QZ, these
surveys produced relatively few useful quasar groups because of the reduced
selection efficiency at $z\ga 2.5$ combined with a bright magnitude limit. In
fact, most well-studied close groups of quasars at $z>2$ were found by slitless
spectroscopy. Follow-up spectroscopy of candidates by \citet{bohuski79} revealed
13 $z>1.5$ quasars on $2.1$~deg$^2$ \citep{jakobsen92} with two showing
correlated complex intergalactic \ion{C}{iv} absorption at $1.48<z<2.15$,
indicative of an elongated superstructure extending over $17\farcm 9$ on the sky
\citep[e.g.][]{jakobsen86,dinshaw96}. Another group discovered by
\citet{sramek78} contains now 6 QSOs at $2.49\le z\le 3.45$ within a radius of
20\arcmin\ around Q~1623$+$2653, all of which are bright enough for correlation
studies of the IGM \citep[e.g.][]{crotts98}. \citet{williger96} reported 25
quasars at $1.5\la z\la 3.4$ within a $\sim 1$~deg$^2$ region and used these to
reveal large-scale structure in the IGM \citep{williger00,liske00b}. At lower
redshifts, slitless surveys revealed large associations of quasars at similar
redshift \citep{crampton90,clowes99}.

Groups of quasars, or more generally speaking, active galactic nuclei, have also
been discovered by recent deep multi-wavelength surveys in selected fields, such
as in the Chandra Deep Fields North \citep{barger03,cowie04} and South
\citep{szokoly04,wolf04}, the Marano field \citep{zitelli92,krumpe07}, or the
COSMOS field \citep{prescott06,trump07}. However, the majority of the AGN thus
found are too faint for follow-up studies at a spectral resolution allowing for
meaningful IGM studies.

Complementary to slitless spectroscopy, quasar pairs can be found from
similarities in multi-colour space. Using this approach, \citet{hennawi06b}
confirmed 40 new associated and 73 projected quasar pairs with separations
$\vartheta<1\arcmin$ from a sample of faint $i\la 21$ quasar pair candidates
selected via SDSS photometry. Follow-up spectra revealed transverse clustering
of optically thick absorption systems near foreground quasars
\citep{hennawi06a,hennawi07}.

Targeted deep surveys of sky regions around well-studied high-redshift quasars
are rare. An exception are many of the fields selected for the Lyman-break
galaxy survey by \citet{steidel03} which were centred on bright quasars to
correlate the galaxies with the intergalactic absorption along the sightline
\citep{adelberger05}. 
Here we describe a systematic search for apparent quasar groups at $1.7\la
z\la 3.6$ in the southern hemisphere, targeting fields centred on known bright
$z>2.7$ quasars that had been observed at high resolution with the UV-Visual
Echelle Spectrograph (UVES) at the VLT.  We already reported results for two
special fields \citep[][hereafter Papers~I and II,
respectively]{worseck06,worseck07}. The present paper is devoted to present
the entire survey. In Sect.~\ref{qnq_surveyobs} we describe the slitless
spectroscopic survey observations. Section~\ref{qnq_surveyred} outlines the
automatic reduction pipeline developed for these data.
Section~\ref{qnq_candselection} describes the semi-automatic selection of
quasar candidates. We report on the follow-up slit spectroscopy of candidates
in Sect.~\ref{qnq_followup}, followed by a brief discussion of the properties
of confirmed quasars and the remaining candidates (Sect.~\ref{qnq_results}).
In Sect.~\ref{qnq_discussion} we present the resulting quasar groups and
discuss the efficiency and completeness of our survey.  We conclude in
Sect.~\ref{qnq_conclusions}.

\begin{figure}
\centering
\includegraphics[scale=0.80]{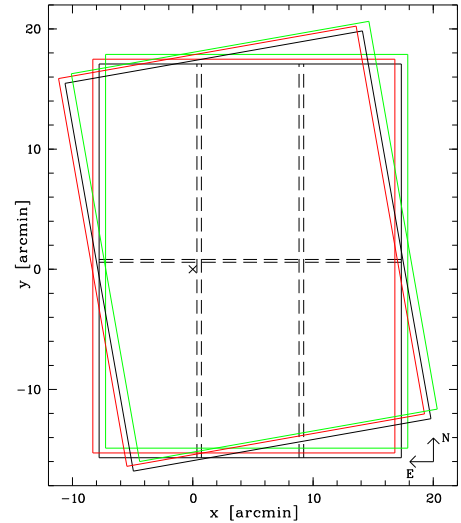}
\caption{\label{qnq_wfidither} Schematic view of the WFI dither pattern. The
full lines denote the edges of the $25\farcm 1\times 32\farcm 7$ field of view
in the dithered 600~s exposures. Dashed lines mark the 6 2k$\times$4k chips with
the inter-chip gaps. After rotating the instrument by $10\degr$ about its origin
in the focal plane (cross) three further 600~s exposures were taken.}
\end{figure}

\begin{table*}
\caption{Observing log of the slitless survey observations. The columns list the
quasar at the field centre with redshift and celestial coordinates, the night of
observation, the employed filters, instrument rotations, total exposure time per
band and the seeing.}

\label{qnq_wfiobservinglog}
\centering\scriptsize
\begin{tabular}[t]{lcccccccc}\hline\hline\noalign{\smallskip}
Field		&$z$	&$\alpha$~(J2000) 		&$\delta$~(J2000) 		&Night		&Filters&Rotations~[\degr]&Exposure~[s]&Seeing~[\arcsec]\\\noalign{\smallskip}\hline\noalign{\smallskip}
\object{Q~0000$-$263}	&$4.125$&00$^{\mathrm{h}}$03$^{\mathrm{m}}$22\fs91	&$-$26\degr03\arcmin16\farcs8	&03~Oct~2002	&$B$, $V$&0	&1800	&0.7--1.0\\
		&	&						&					&04~Oct~2002	&$B$, $V$&10	&1800	&0.9--1.4\\
\object{Q~0002$-$422}	&$2.767$&00$^{\mathrm{h}}$04$^{\mathrm{m}}$48\fs11	&$-$41\degr57\arcmin28\farcs8	&01~Oct~2002	&$B$, $V$&0, 10		&3600	&1.3--1.6\\
\object{Q~0055$-$269}	&$3.665$&00$^{\mathrm{h}}$57$^{\mathrm{m}}$57\fs92	&$-$26\degr43\arcmin14\farcs2	&02~Oct~2002	&$B$, $V$&0, 10		&3600	&0.9--1.5\\
\object{Q~0302$-$003}	&$3.285$&03$^{\mathrm{h}}$04$^{\mathrm{m}}$49\fs86	&$-$00\degr08\arcmin13\farcs4	&03~Oct~2002	&$B$, $V$&0	&1800	&0.9\\
		&	&						&					&03~Oct~2002	&$B$	&10	&1800	&0.9\\
		&	&						&					&04~Oct~2002	&$V$	&10	&1800	&0.9--1.1\\
\object{Q~0347$-$383}	&$3.220$&03$^{\mathrm{h}}$49$^{\mathrm{m}}$43\fs68	&$-$38\degr10\arcmin31\farcs3	&27~Feb~2003	&$B$, $V$&10	&1800	&1.0\\
		&	&						&					&27~Feb~2003	&$V$	&0	&1800	&1.0\\
		&	&						&					&28~Feb~2003	&$B$	&0	&1800	&1.9\\
\object{CTQ~0247}	&$3.025$&04$^{\mathrm{h}}$07$^{\mathrm{m}}$17\fs99	&$-$44\degr10\arcmin13\farcs4	&30~Sep~2002	&$B$, $V$&0	&1800	&1.3--1.8\\
		&	&						&					&01~Oct~2002	&$B$, $V$&10	&1800	&1.4--1.8\\
\object{Q~0420$-$388}	&$3.120$&04$^{\mathrm{h}}$22$^{\mathrm{m}}$14\fs81	&$-$38\degr44\arcmin52\farcs9	&26~Feb~2003	&$B$, $V$&0, 10	&3600	&0.8\\
\object{PKS~0528$-$250} &$2.813$&05$^{\mathrm{h}}$30$^{\mathrm{m}}$07\fs96	&$-$25\degr03\arcmin29\farcs9	&03~Nov~2002	&$V$	&10	&1800	&1.1\\
		&	&						&					&04~Nov~2002	&$V$	&0	&1800	&1.2\\
		&	&						&					&28~Feb~2003	&$B$	&0, 10	&3600	&1.2\\
\object{HE~0940$-$1050} &$3.088$&09$^{\mathrm{h}}$42$^{\mathrm{m}}$53\fs40	&$-$11\degr04\arcmin25\farcs0	&26~Feb~2003	&$B$, $V$&0, 10	&3600	&0.8\\
\object{CTQ~0460}	&$3.139$&10$^{\mathrm{h}}$39$^{\mathrm{m}}$09\fs51	&$-$23\degr13\arcmin25\farcs7	&27~Feb~2003	&$B$, $V$&0, 10	&3600	&1.0--1.5\\
\object{BR~1117$-$1329}	&$3.958$&11$^{\mathrm{h}}$20$^{\mathrm{m}}$10\fs30	&$-$13\degr46\arcmin25\farcs0	&28~Feb~2003	&$B$, $V$&0, 10	&3600	&$>2$\\
\object{BR~1202$-$0725}	&$4.690$&12$^{\mathrm{h}}$05$^{\mathrm{m}}$23\fs12	&$-$07\degr42\arcmin32\farcs5	&26~Feb~2003	&$B$, $V$&0	&1800	&0.8\\
		&	&						&					&27~Feb~2003	&$V$	&10	&1800	&1.6\\
\object{Q~1209$+$093}	&$3.291$&12$^{\mathrm{h}}$11$^{\mathrm{m}}$34\fs95	&$+$09\degr02\arcmin20\farcs9	&27~Feb~2003	&$B$, $V$&0, 10	&3600	&1.6\\
\object{Q~1451$+$123}	&$3.246$&14$^{\mathrm{h}}$54$^{\mathrm{m}}$18\fs61	&$+$12\degr10\arcmin54\farcs8	&28~Feb~2003	&$B$, $V$&0, 10	&3600	&0.8--2.0\\
\object{PKS~2126$-$158}	&$3.285$&21$^{\mathrm{h}}$29$^{\mathrm{m}}$12\fs18	&$-$15\degr38\arcmin41\farcs0	&30~Sep~2002	&$B$, $V$&0, 10	&3600	&0.8--1.0\\
\object{Q~2139$-$4434}	&$3.214$&21$^{\mathrm{h}}$42$^{\mathrm{m}}$25\fs81	&$-$44\degr20\arcmin17\farcs2	&03~Oct~2002	&$B$, $V$&0, 10	&3600	&1.0--1.5\\
\object{HE~2243$-$6031}	&$3.010$&22$^{\mathrm{h}}$47$^{\mathrm{m}}$09\fs10	&$-$60\degr15\arcmin45\farcs0	&02~Oct~2002	&$B$, $V$&0, 10	&3600	&1.0--1.7\\
\object{HE~2347$-$4342}	&$2.885$&23$^{\mathrm{h}}$50$^{\mathrm{m}}$34\fs21	&$-$43\degr25\arcmin59\farcs6	&04~Oct~2002	&$B$, $V$&0, 10	&3600	&0.8--1.2\\
\noalign{\smallskip}\hline
\end{tabular}
\end{table*}

\section{Survey observations}
\label{qnq_surveyobs}

The survey was carried out using the ESO Wide Field Imager
\citep[WFI,][]{baade99} at the ESO/MPI $2.2$~m Telescope (La Silla) in its
slitless spectroscopic mode \citep{wisotzki01}. Since this mode has been
rarely used we shortly describe its main characteristics.

The WFI is a focal-reducer type camera offering a field of view of
$34\arcmin\times 33\arcmin$ sampled by a mosaic of 8 2k$\times$4k CCDs with
$0\farcs 238$/pixel. In the slitless spectroscopic mode a low-resolution grism
is placed in the converging beam of the telescope in front of the WFI in order
to disperse the light of every object in the field of view. A blue-blazed
grism and a red-blazed grism are available. However, the red-blazed grism
(R50, dispersion $\sim 7$\AA/pixel, blaze wavelength 6000\AA,
$\Delta\lambda\sim 50$\AA\ FWHM) has a much higher throughput for 1st-order
spectroscopy even in the blue \citep{wisotzki01}, rendering the blue-blazed
B50 grism almost obsolete. We only used the R50 grism. As both grisms were
originally constructed for the prime focus of the ESO 3.6~m telescope (long
decommissioned) and only retrofitted into WFI, the size of the grisms does not
fully match that of the WFI field of view. In a $7\farcm 2$ strip on the left 
side of the chip mosaic the light passes undispersed, effectively reducing
the usable field of view to 6 of the 8 WFI CCDs, or
$25\arcmin\times 33\arcmin$.

During two visitor mode runs in October 2002 (5 nights) and February 2003 (3
nights) we observed in total 18 fields centred on bright high-redshift
quasars with available UVES spectra in order to find faint quasars in their
surroundings. We additionally included the Extended Chandra Deep Field South
(ECDFS) to compare our selection with the results of deeper multiwavelength
surveys.

The sky transparency during the October 2002 run was variable with some thick
cirrus clouds passing occasionally, but mostly the sky was clear. Conditions
were clear to photometric in the February 2003 run. The seeing varied over a
broad range of values (see Table~\ref{qnq_wfiobservinglog}) but was mostly of
the order of 1\arcsec. Since the spectral resolution in slitless spectroscopy
is given by the seeing, this quantity varied by $\lambda/\Delta\lambda=30$--50.
We used the broadband $B$ and $V$ filters of the WFI to lower the effectively
undispersed sky background and to reduce the degree of crowding by limiting
the length of the spectra. Crowding was further accounted for by taking the
spectra in two instrument rotations (0\degr~and 10\degr). For all fields
(except the one centred on BR~1202$-$0725) three dithered 600~s exposures were
taken per band and rotation, resulting in a total exposure time of 1~h per
band. Figure~\ref{qnq_wfidither} illustrates the observing pattern. The
dithered exposures produced a contiguous $\sim 26\farcm 2\times 33\farcm 5$
field of view per rotation angle. The combination of the R50 grism with the
broadband $B$ and $V$ filters resulted in a spectral coverage from the blue
sensitivity cutoff at $\sim 4200$~\AA\ to 5800~\AA. An unfiltered spectrum of
the low-redshift emission-line galaxy HE~1250$-$0256 provided the input for an
approximate wavelength calibration of the slitless data. The
spectrophotometric standard stars HD~49798, LTT~7987 and GD~108 were observed
in astronomical twilight for relative flux calibration. Blank sky fields
observed during twilight provided flat fields. The $V$ band exposures
of the field centred on PKS~0528$-$250 were retrieved from the ESO Science
Archive (PI L.~Vanzi). Table~\ref{qnq_wfiobservinglog} summarises the slitless
survey observations.

The nominal surveyed area was $\simeq 4.39$~deg$^2$. However, due to the two
rotation angles of the instrument, the field edges received only $\sim 1/2$ of
the exposure time. Moreover, spectra located near interchip gaps are affected by
dithering. The net exposure time per field is further decreased for some
objects due to contamination by nearby other sources or spectral orders of
bright stars (see Fig.~\ref{qnq_maskplot} below). However, the two instrument
rotations ensured that a clean spectrum of almost every object was obtained.

Direct images are necessary for object identification in the slitless data. We
primarily relied on images from the Digitised Sky Survey
(DSS)\footnote{%
  The Digitized Sky Surveys were produced at the Space Telescope Science Institute
  under U.S. Government grant NAG W-2166. The images of these surveys are based on
  photographic data obtained using the Oschin Schmidt Telescope on Palomar Mountain
  and the UK Schmidt Telescope. The plates were processed into the present
  compressed digital form with the permission of these institutions.}
available for all our fields. For three fields we also employed direct WFI
$BVR$ images taken in service mode.

\begin{figure*}
\sidecaption
\includegraphics[width=12cm]{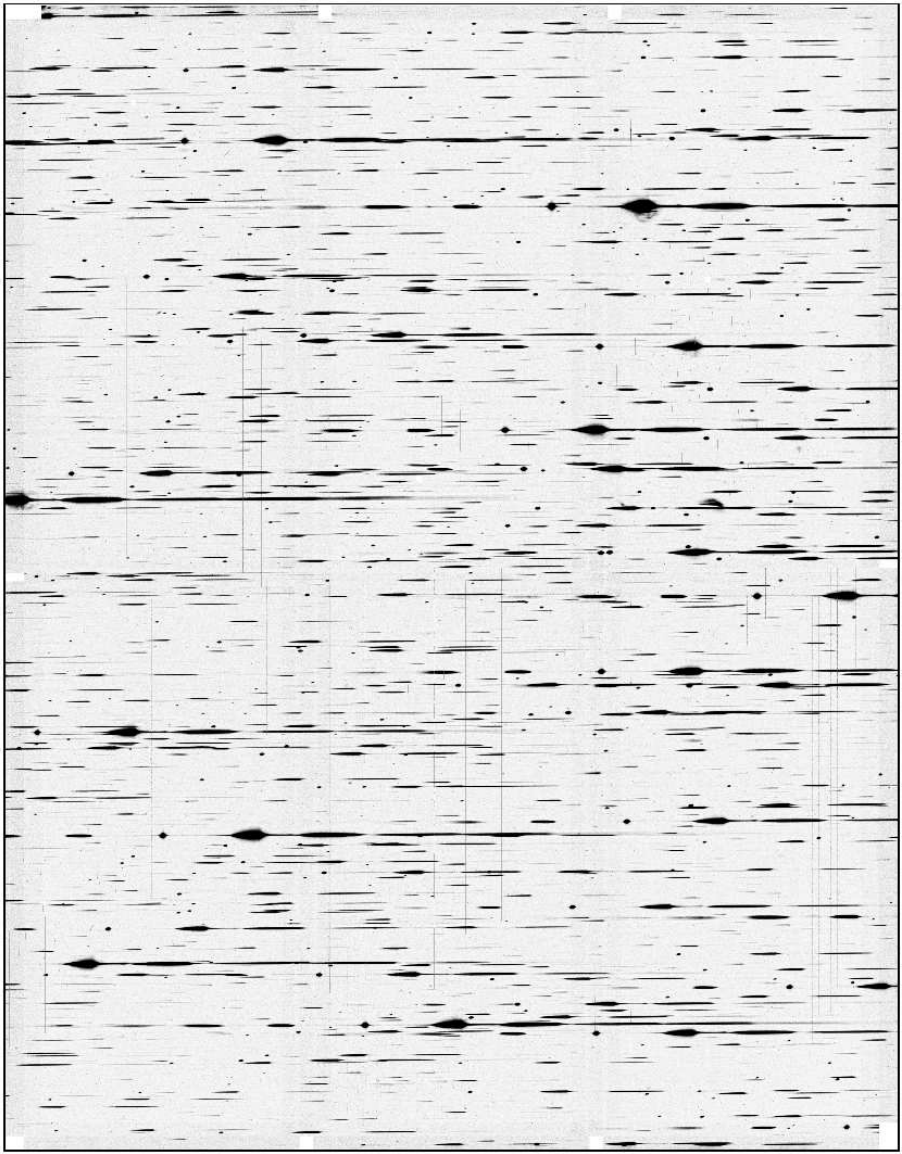}
\caption{\label{he0940b0_bw2} Flat-fielded sky-subtracted and combined slitless
spectroscopic WFI field in the vicinity of HE~0940$-$1050. The spectra were
taken with the $B$ filter and without rotation (i.e.\ north is up and east is
left on the image). The dispersion runs horizontally from left to right. The
three dithered 600~s exposures were combined to fill the inter-chip gaps. The
(linear) intensity of each pixel is weighted by its effective exposure time.
The total dimension of the combined image is $26\farcm 16\times 33\farcm 53$.}
\end{figure*}

\section{Survey data reduction}
\label{qnq_surveyred}

\subsection{Reduction steps}

The slitless data were reduced with a semi-automatic pipeline running under 
ESO-MIDAS, but largely consisting of our own dedicated software modules.
For several parts of the data reduction we could use the extensive toolbox
developed for the slitless spectroscopy reduction pipeline of the Hamburg/ESO
Survey \citep[HES,][]{wisotzki00}. This was supplemented by new software where
necessary. Each exposure was reduced separately before combining the extracted
spectra. Briefly, the data reduction comprised the following steps:

\begin{enumerate}
\item\textit{Bias subtraction:} 
Bias subtraction was performed by taking the median of the overscan regions of
each WFI chip and subtracting it from the science exposure. This procedure
assured that always the correct bias value was subtracted, since restarts of
the electronics were necessary a few times during the nights, altering the
bias level.

\item\textit{Flat-fielding:}
The flat-fielding of slitless spectroscopic data is complicated by the fact that
in a science exposure a pixel of a slitless spectrum receives the effectively
undispersed broadband sky background and nearly monochromatic light from the
object, whereas it is exposed only to the undispersed broadband skylight when
obtaining the twilight flat field. After some experiments, we obtained good
master flat fields by normalising the individual flat field frames by a smoothed
3rd-order polynomial fit along the dispersion direction followed by averaging
the normalised images. Flat-fielding of the science data was performed by
dividing by the appropriate master flat field. 

\item\textit{Sky subtraction:}
The sky background was subtracted using a background image created after masking
the spectra using a programme developed for HES. First, the mode of the image
was estimated in coarse cells before masking the regions containing spectra by
a $\sigma$-clipping algorithm. Subsequently, the average of the unmasked pixels
was computed on a finer grid and the variable background was determined by
bilinear interpolation. The parameters for creating the masks were varied until
the sky-subtracted image was free from artefacts due to over- or
undersubtraction near the spectra of bright objects, and the sky residuals were
consistent with zero.

\item\textit{Astrometric transformation:}
Bright stars on the DSS images were used for an iterative astrometric
transformation that yielded the coordinates for the extraction of the spectra.
In an initial step, several objects in each field were visually identified
in both direct and spectral images. The cutoff of the object flux at the red
end of both filters yielded a well-defined fixpoint for all considered
sources. Essentially this was the only reduction step requiring human
intervention prior to the selection of candidates. The identified stars
defined a preliminary bilinear transformation from the sky to spectral plate 
coordinates, which was then refined by including several
more stars automatically identified in the direct data. Rejecting outliers,
the final sets of transformation coefficients were based on $\sim 15$--50
sources per detector chip subfield and provided a mapping between DSS 
coordinates and nominal red cutoff of the 1st-order spectral positions with 
a typical accuracy of $\pm 2$ pixels (rms).

\item\textit{Input source catalogue:}
The object search routine was used to generate deep source lists of the DSS
images. We required an object to have at least 3 pixels $4\sigma$ above the
background level. The lists also contained objects located at the field edges
that were not recorded in every exposure due to dithering or instrument
rotation. The corresponding coordinates of the sources on the slitless images
were derived from the astrometric transformation.
For most of the fields the limiting magnitude of the source catalogues was
$V_\mathrm{lim}\simeq 21$ (Sect~\ref{qnq_surveydepth}).

\begin{figure*}
\centering
\includegraphics[scale=1]{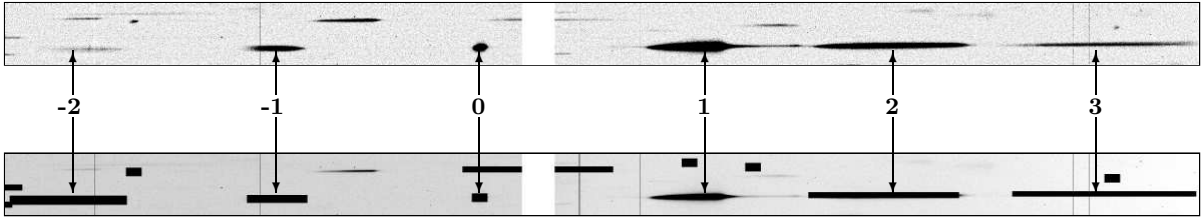}
\caption{\label{qnq_maskplot} Masking of unwanted spectral orders. The upper
panel shows a $850\arcsec\times 45\arcsec$ region of a sky-subtracted 600~s $B$
band exposure of the field around HE~0940$-$1050. A bright star in this region
generates the indicated different spectral orders in the slitless exposure that
may contaminate 1st-order spectra of nearby sources. The white stripe marks a
vertical inter-chip gap. The lower panel displays the corresponding region of
the variance image. Black rectangles show masked spectral orders of the bright
star and other objects.
Dark vertical stripes correspond to masked bad columns on the chips.}
\end{figure*}

\item\textit{Masking of spectral orders and defects:} In contrast to an
objective prism, a grism produces several spectral orders.
Figure~\ref{he0940b0_bw2} shows a combined $26\farcm 2\times 33\farcm 5$
slitless spectroscopic WFI image centred on the quasar HE~0940$-$1050 taken in
the $B$ filter. More than 60\% of the flux is concentrated in the +1st order
spectra, with most of the remaining flux going into the 2nd and 0th orders
\citep{wisotzki01}. Thus, for faint objects only these orders are easily
visible in the data. Inevitably, brighter stars produce also higher spectral
orders (and negative ones, streching to the left in
Fig.~\ref{he0940b0_bw2}). In all cases, overlaps between different orders of
different objects is in danger of producing undesirable artefacts; a
particular nuisance for our programme are the blob-like zero-order images that
superposed on other spectra will very much resemble emission lines. We
developed a masking scheme combined with the extraction (next step) where all
pixels affected by unwanted spectral orders from another object (depending on
the magnitude of the recorded object) were flagged in the corresponding
variance frame. Figure~\ref{qnq_maskplot} shows a small section from a
spectral image together with its corresponding variance frame where all
spectral orders of an object contaminating other spectra were masked. Also
included in the masks were all detector bad pixels and columns as well as
cosmic ray hits.

\item\textit{Optimal extraction:} The slitless spectra of the sources in the
input catalogues of each exposure were extracted assuming a Gaussian spatial
profile with predefined centroid and full width at half maximum (FWHM). We
found the bright stars used for the coordinate transformation to be reliable
tracers of the centroid of a spectrum and its FWHM. The centroid vector is
usually slightly tilted with respect to the detector columns due to
imperfect alignment of the dispersion axis with the CCD rows. The FWHM is
primarily set by the seeing, but the grism optics make it increase slightly
from east to west over the field of view. All objects in the input catalogue
were then extracted with the pre-determined centroid and FWHM vectors of each
exposure, maximising the signal-to-noise ratio ($S/N$) of faint objects. Along
with each spectrum, a $1\sigma$ noise array was extracted. The extracted
spectra of all 600~s exposures were adjusted to a common counts scale to
correct for possible sky transparency variations, before averaging them by
weighting with their inverse pixel variances. The masking usually (but not
always) ensured that overlaps with other orders were adequately reflected in
the pixel variances, so that affected objects were generally free of overlap
artefacts after averaging the spectra taken at the two rotator angles.

\item\textit{Wavelength calibration:} Gaussian fits to emission lines of the
high-$S/N$ spectrum of the low-redshift emission line galaxy HE~1250$-$0256
yielded an approximate wavelength scale with a linear dispersion of $\simeq
6.7$\AA/pixel. Both the dispersion and the wavelength zero point (as
determined by the astrometric transformation) change somewhat over the field 
of view due to optical distortions induced by the grism \citep{wisotzki01}.
These variations ultimately limited the accuracy by which wavelengths could be
measured in any given spectrum to roughly $\Delta\lambda \simeq \pm 15$~\AA.

\item\textit{Flux calibration:}
Observations of three spectrophotometic standard stars yielded the instrument
sensitivity curves of the spectra taken in the two filters
(Fig.~\ref{qnq_filtercurves}). Flux calibration of the spectra was achieved by
dividing the extracted spectra by the appropriate sensitivity curve scaled to
the nominal exposure time of 600~s. Because of changes in sky transparency
during the first observing run and because of the long intervals between 
standard star observations, the flux scale should be considered as relative, 
although we found good agreement between the published magnitudes and the
integrated slitless spectroscopic magnitudes of the known quasars in the fields.

\item\textit{Final spectra:}
The calibrated filtered spectra were spliced together in their overlapping
spectral range. Spectra taken in both instrument rotations were reduced
separately and then combined by weighting with their inverse variances. The
final stack of slitless spectra from a field contained every source from
the DSS input catalogue that had been recorded even in a single exposure due
to dithering and field rotation.
\end{enumerate}

Compared to standard imaging with the WFI, the inserted red-blazed R50 grism
reduces the instrument throughput by $\sim 15\%$ and $\sim 36\%$ in the $V$
and $B$ bands, respectively. The low sensitivity in the blue is to some extent 
balanced by a lower sky background compared to the $V$ band.
After inspecting the extracted spectra of one field we defined the useful
wavelength range to 4200~\AA $\le$ $\lambda$ $\le$ 5800~\AA. 
In each field we extracted 800--1600 meaningful spectra, the exact number
varying with the depth of the input catalogue and with Galactic latitude.
In total, slitless spectra of $\sim 29000$ objects were extracted in the 18
survey fields. The control field in the ECDFS contained an additional 1000
sources.

\begin{figure}
\resizebox{\hsize}{!}{\includegraphics{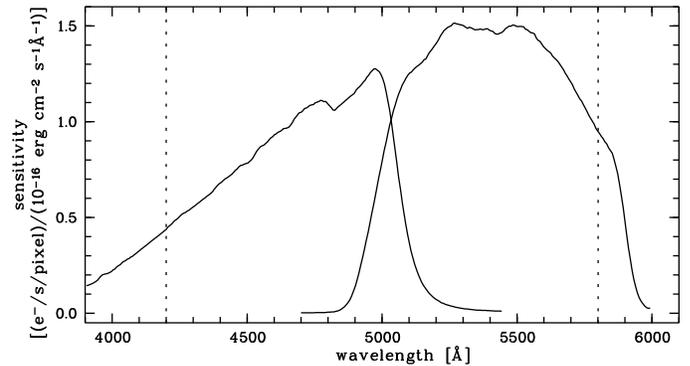}}
\caption{\label{qnq_filtercurves} Sensitivity curves of the WFI $B$ and $V$
filters derived from the spectrophotometric standard HD~49798. At
$\sim 4700$~\AA\ one triggered photoelectron per second per $\simeq 6.7$~\AA\
pixel corresponds to a flux density of
$10^{-16}\,\mathrm{erg}\,\mathrm{cm}^{-2}\,\mathrm{s}^{-1}\,\mathrm{\AA}^{-1}$.
Dotted lines mark the considered wavelength range.}
\end{figure}

\subsection{Survey depth}
\label{qnq_surveydepth}

\begin{figure}
\resizebox{\hsize}{!}{\includegraphics{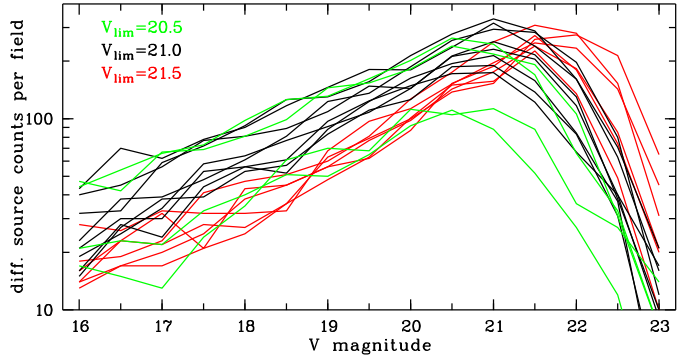}}
\caption{\label{qnq_diffobjcounts} Differential source density of the survey
fields in $\Delta V=0.5$~mag bins derived from the DSS plates. The colours
indicate different estimated limiting magnitudes
($V_\mathrm{lim}\simeq 20.5$: CTQ~0460, Q~0302$-$003, Q~1209$+$093, PKS~2126$-$158,
$V_\mathrm{lim}\simeq 21.0$: BR~1117$-$1329, BR~1202$-$0725, HE~0940$-$1050,
HE~2243$-$6031, PKS~0528$-$250, Q~0420$-$388, Q~1451$+$123, Q~2139$-$4434,
$V_\mathrm{lim}\simeq 21.5$: CTQ~0247, Q~0055$-$269, HE~2347$-$4342, Q~0000$-$263,
Q~0002$-$422, Q~0347$-$383).}
\end{figure}

The magnitude limit of our survey obviously depends on both the depth of the
input source catalogues and on the obtainable S/N in the slitless spectroscopy
data. We estimated the completeness of the source catalogues as a function of
magnitude by considering the differential source counts.
Figure~\ref{qnq_diffobjcounts} shows the differential surface densities per
$\Delta V=0.5$~mag obtained from the DSS source catalogues in the 18 fields.
From the breaks in the differential number counts we conclude that the input
catalogues are reasonably complete to $V\sim 21$, with 4 fields being shallower
by $\sim 0.5$~mag. Similar limiting magnitudes were obtained in the three fields
with direct WFI data.

\begin{figure}
\resizebox{\hsize}{!}{\includegraphics[angle=270]{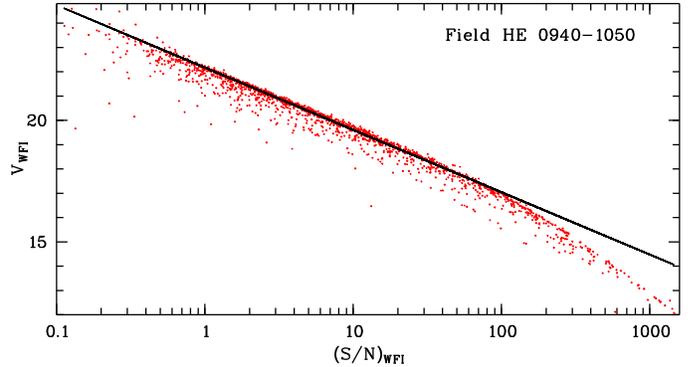}}
\caption{\label{qnq_snmag} Integrated WFI $V$ magnitude vs.\ median $S/N$ in
the WFI spectra in the field of HE~0940$-$1050 (dots). The line shows the
fitted log-linear relation of the magnitude and the $S/N$ for this field.}
\end{figure}

The limiting magnitude of the slitless survey material was determined by
evaluating the continuum $S/N$ of the extracted spectra. The $V$ magnitude of
each object was calculated by integrating the flux over the $V$ band which is
completely covered by the spectra. We estimate that the $V$ magnitudes of
$V\la 20$ objects are accurate to $\sim 0.1$~mag. Fainter sources have larger 
uncertainties due to their lower S/N. The S/N is a function of the
wavelength-dependent system sensitivity (Fig.~\ref{qnq_filtercurves}) and of
the source spectral energy distribution. We defined the median S/N in the $V$ 
band as a characteristic value. Recall that the extraction yielded a noise
array along with each spectrum, so that the S/N can be estimated rather
accurately. In Fig.~\ref{qnq_snmag} we plot the median 
S/N of all extracted $\sim 1800$ spectra in one particular field,
as a function of $V$ magnitude. One can see that a S/N of 3 is obtained
at $V\simeq 21.0$, and for $V\simeq 22.0$ a S/N of unity is reached.

Some objects appear to have too low S/N for their $V$ magnitude; these
generally have shorter net exposure times induced by dithering and field
rotation. After excluding these outliers we fitted a simple log-linear 
relation for the S/N as a function of magnitude,
\begin{equation}\label{qnqeq_vsn}
V_\mathrm{WFI}=-2.5\log{\left(S/N\right)}+22.05
\end{equation}
with a root-mean-square deviation between individual fields of $\sim 0.1$~mag.
The $S/N$ of very bright objects is systematically lower than this relation
because of large non-Gaussian tails of the PSF which were not taken into
account in the extraction, and because of detector saturation occurring at
$V_\mathrm{WFI}\la 13.4$ (corresponding to 130000~$\mathrm{e}^-$ per pixel in
600~s).

We conclude that at a magnitude of $V\simeq 21$, our survey should be
reasonably complete in terms of the input catalogues, and that the slitless
spectra at these magnitudes are generally still good enough to search for
emission-line objects. We demonstrate below that the selection of even fainter
$z>2.4$ quasars with prominent Ly$\alpha$ lines was also possible, in some
fields down to $V\simeq 22$ or beyond (corresponding to a continuum S/N of 1
in the slitless spectra).  However, it is clear that only the most promising
candidates will be selected at these faint levels. Moreover, we have seen that
at $V\simeq 22$, already the input catalogues are heavily affected by
approaching to the limits of the DSS itself. We return to a more quantitative
assessment of (in)completeness of the quasar samples in
Sect.~\ref{qnq_completeness} below.

\begin{figure}
\resizebox{\hsize}{!}{\includegraphics{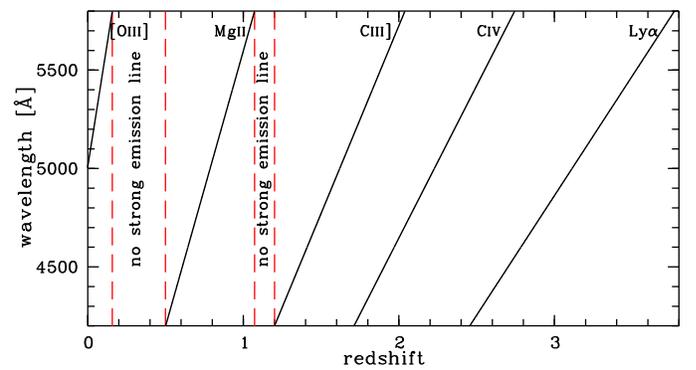}}
\caption{\label{qnq_lineshift} Visibility of strong emission lines over the
covered wavelength range. Dashed lines mark gaps without major emission lines
where quasar candidates are missed.}
\end{figure}

\section{Selection of quasar candidates}
\label{qnq_candselection}

\subsection{Selection criteria}
\label{qnq_selectcrit}

Quasar candidates were semi-automatically selected from the calibrated slitless
spectra on the basis of prominent emission lines. Figure~\ref{qnq_lineshift}
shows the redshift ranges in which major quasar emission lines can be observed
in the slitless spectra. The detectability of the various emission lines depends
on their equivalent widths and the $S/N$ in the data
(Sect.~\ref{qnq_completeness}). Moreover, quasar candidates could not be selected
in two narrow low-redshift ranges due to the lack of major emission lines
(see Fig.~\ref{qnq_lineshift}).

We first estimated a (pseudo-)continuum for each spectrum by applying a simple
median filter with a window width of 400~\AA. The
continuum-subtracted spectra were searched automatically for emission line
objects by cross-correlating them with Gaussian emission line templates. The
fields around Q~0000$-$263, Q~0302$-$003 and Q~2139$-$4434 were chosen for
testing the continuum subtraction and the line detection algorithm, ensuring
that all emission line objects selected by eye were also found automatically.
Finally, we chose two Gaussians with dispersions $\sigma_\lambda=40$~\AA\ and
of $\sigma_\lambda=70$~\AA\ in order to match emission lines with small and
large widths, respectively. Only features detected with a total line S/N $\ge$ 3
and an observed equivalent width of $W_\mathrm{obs}\ge 20$~\AA\ were kept.
 
At our spectral resolution, the continuum subtracted spectra of normal stars
often appeared to show `emission lines' that in reality were artefacts of poor
continuum estimation and oversubtraction. The initial samples of emission line
objects contained therefore $\sim 1/4$ of all extracted spectra. However, it
was quite straightforward to beat down this number significantly. First of
all, the equivalent widths of these features were generally quite small. 
Secondly, most of the spuriously selected stars had rather red colours.
We quantified a colour measure by defining a `hardness ratio'\footnote{%
  We borrowed this concept from the X-ray community because it is more
  reliable and less prone to noise peaks for very low S/N data.}
as
\begin{equation}
H=\frac{f_{\lambda,V}-f_{\lambda,B}}{f_{\lambda,V}+f_{\lambda,B}}
\end{equation}
where $f_{\lambda,B}$ and $f_{\lambda,V}$ are the average fluxes in the
intervals 4200~\AA $\le$ $\lambda$ $\le$ 5000~\AA\ 
and 5000~\AA $\le$ $\lambda$ $\le$ 5800~\AA, respectively.
Thus, the larger $H$, the redder the object.

We finally decided to treat emission line candidates with large equivalent
widths differently from those with weak line detections. A strong-lined
object was defined as one with a line S/N $\ge$ 3 and $W_\mathrm{obs}\ge
100$~\AA. If the line was detected in the $B$ band, the object needed to be
moderately blue, $H-\sigma_H\le 0.046$; this criterion removed most of the
red stars. If the line was detected in the $V$ band, no further restriction
was applied.
A line detection with small equivalent width, $20 \le W_\mathrm{obs} < 100$~\AA,
needed always to be accompanied by a blue but meaningful spectrum, given by the
conditions $H+\sigma_H<0.15$ and $H-\sigma_H>-1$.

Essentially all Ly$\alpha$ detected quasars, but also some of the \ion{C}{iv}
and [\ion{O}{iii}] detections have large equivalent widths and therefore fell
into the first set of criteria.  Quasars at lower redshifts generally have
weaker lines (\ion{C}{iv}, \ion{C}{iii}], \ion{Mg}{ii} and [\ion{O}{iii}]),
but are distinctly blue. Note that our highly conservative colour cuts were
only applied \emph{after} the emission line object selection; these cuts would
by no means have sufficed to select quasar candidates on the basis of their
colours.

\begin{figure}
\resizebox{\hsize}{!}{\includegraphics{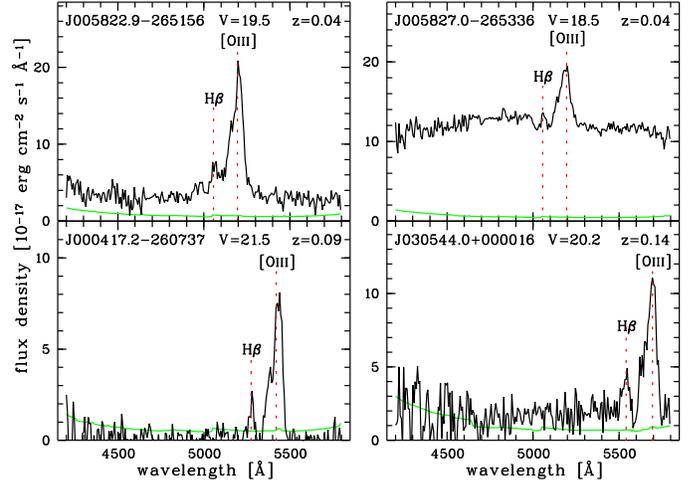}}
\caption{\label{qnq_emlgal} Slitless spectra (black) and corresponding $1\sigma$
noise arrays (green/grey) of four emission line galaxies showing blended
[\ion{O}{iii}] emission. H$\beta$ can be recognised as well.}
\end{figure}

After these cuts we visually inspected the selected objects to cull the quasar
candidates from the remaining sample of slitless spectra, dominated by three
types of spurious detections: (i) M stars with strong TiO absorption bands
that our selection criteria did not catch; these features are always at the
same wavelengths and could be recognised with good confidence after some
training. (ii) Low-redshift emission line galaxies with [\ion{O}{iii}] visible
at $z\la0.15$; at the low resolution of the WFI spectra, the [\ion{O}{iii}]
doublet is blended, resulting in an asymmetric single line that could be
unambiguously identified in almost all cases (see Fig.~\ref{qnq_emlgal} for
examples). Since emission line galaxies are interesting in their own right, we
kept them in a separate bin. (iii) Defective spectra were overlapping 1st
order spectra of two close objects, or spectra with unrecognised zero order
contamination or spectra with unphysical breaks at $\sim 5000$~\AA\ resulting
from an imperfect splicing of the two bandpasses. Any remaining spectral
overlaps were easily identified by comparing the spectra obtained at the two
instrument rotation angles. All emission line candidates that passed
the visual check of the one-dimensional spectra were finally verified on the
slitless WFI images to check for any remaining flaws that could result in a
spurious emission line.  For objects at the field edges detected in only one
instrument rotation (Fig.~\ref{qnq_wfidither}) this visual examination was
essential. In total, 387 emission line objects were selected from the 
$\sim 29000$ objects detected on the DSS images of the 18 survey fields.

\begin{figure*}
\centering\includegraphics[width=\textwidth]{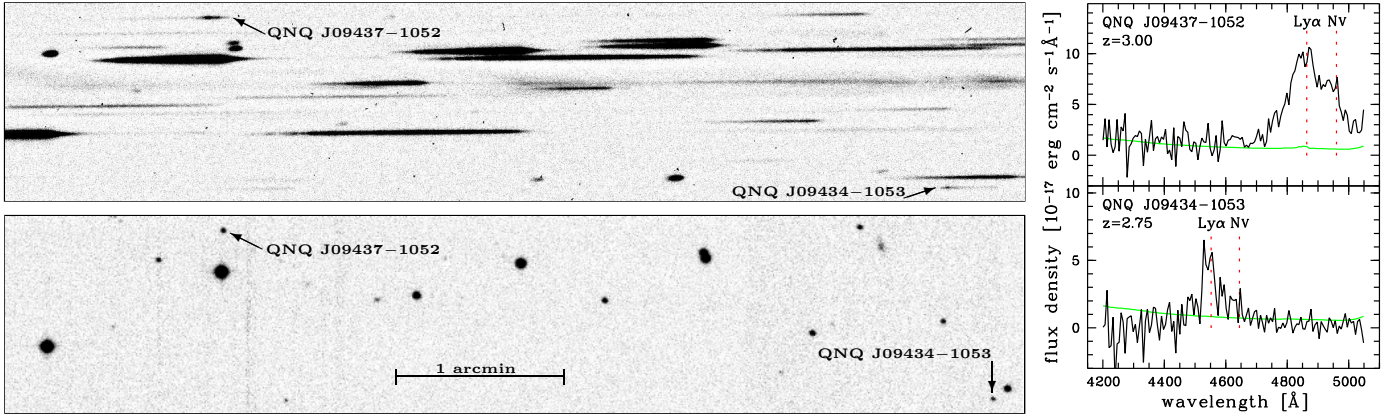}
\caption{\label{qnq_spcdirect} Slitless WFI spectra of two newly discovered
  quasars near HE~0940$-$1050. The upper left panel shows a close-up of the
  combined slitless $B$-filtered exposures in the field of HE~0940$-$1050
  taken without rotation (Fig.~\ref{he0940b0_bw2}). The dispersion runs
  horizontally from left to right. The lower left panel shows a direct WFI
  $B$ band image of the corresponding sky region north-east of HE~0940$-$1050
  ($6\farcm 10\times 1\farcm 19$). The indicated objects display prominent
  emission lines in the slitless data. The right panels show the automatically
  reduced and calibrated one-dimensional $B$ band spectra of the two quasars
  (black) and their $1\sigma$ noise arrays (green/grey). Both of them display
  Ly$\alpha$ emission blended with \ion{N}{v} at the indicated redshifts.}
\end{figure*}

Visual inspection of the sky-subtracted slitless images revealed three
additional candidates that were missed in the DSS catalogues, mainly due to
source blending. These fields were reprocessed with the source catalogues of
available direct WFI data. At the higher resolution of the direct WFI data,
these candidates were resolved and their spectra were successfully extracted.
Figure~\ref{qnq_spcdirect} presents the slitless $B$ band data and the
corresponding direct WFI image of a region near HE~0940$-$1050 where two prime
quasar candidates were discovered. On the DSS image of this field,
QNQ~J09434$-$1053 could not be resolved from the nearby foreground object. The
WFI catalogues did not reveal further undetected candidates, confirming our
expectation that the fraction of candidates lost due to blended sources is very
small.

\subsection{The sample of quasar candidates}

In total, our semi-automated search for emission line objects yielded 390
sources on the surveyed 4.39~deg$^2$. Of those, 38 we rejected as dubious due to
very low $S/N$. The remaining sample of 351 objects consists of 37 known quasars
(including 15 of the 18 targeted central quasars), 169 quasar candidates, and
145 candidate low-redshift star-forming galaxies. The classification was purely
based on the slitless spectra. No morphological selection was made.

Most of the emission line galaxies were identified by their blended
[\ion{O}{iii}] doublet. However, due to low $S/N$ at this low resolution
$R\sim 50$, we could not determine whether the H$\beta$ lines seen in the
spectra are broad or narrow, so that this sample of low-redshift objects might
still contain low-redshift AGN. At the bright end of our survey ($V\la 18$), the
surface density of $z\la 0.15$ emission line galaxies is $\sim 1.1$~deg$^{-2}$,
which compares well with other slitless surveys \citep{salzer02,bongiovanni05}.
As most of the deeper surveys for emission line galaxies select H$\alpha$
emission \citep[e.g.][]{pascual01}, a comparison at fainter limits is difficult.

In total, we rediscovered 37 previously known quasars in the survey fields with
detectable emission lines in the observed wavelength range. Three of the central
quasars (Q~0000$-$263, BR~1117$-$1329 and BR~1202$-$0725) reside at too high
redshift to be detected in emission. Excluding the central bright quasars, 10
out of 12 known quasars are rediscovered in the range of high selection
probability ($V\la 21$, $1.7<z<3.6$, see Sect.~\ref{qnq_completeness} below).
One of the two missed quasars had been recorded in one WFI rotation only,
whereas the other one was not contained in the shallow DSS source catalogue of
the Q~0302$-$003 field (Fig.~\ref{qnq_diffobjcounts}). The redshifts of the
rediscovered quasars measured in the slitless data are consistent
(within $\sigma_z\simeq 0.03$) with the published values
(see Sect.~\ref{qnq_zcomparison} and Fig.~\ref{qnq_wfiqnqcorr} below).

We estimated redshifts of the 169 quasar candidates from the slitless data,
and the candidates were grouped by redshift confidence. The redshifts were
considered to be almost secure if either two emission lines with a consistent
redshift were present in the spectral range, or if a single line could be
confidently identified due to its shape and/or that of the adjacent
continuum. In particular, Ly$\alpha$ emission blended with \ion{N}{v} could
often be unambiguously identified (for examples see
Fig.~\ref{qnq_spcdirect}). Bright low-redshift candidates displayed lines with
small equivalent widths (likely \ion{C}{iii}] or \ion{Mg}{ii}) superimposed on
a blue continuum. Redshift assignments of fainter quasar candidates were less
certain due to the similar equivalent widths of the \ion{C}{iv}, \ion{C}{iii}]
and \ion{Mg}{ii} emission lines. We decided in all doubtful cases to assign
the highest plausible redshifts to the candidates.

For eight of our fields we retrieved WFI$+$grism $R$ band data from the ESO
Science Archive (taken in the course of another programme, PI L.~Vanzi). We
reduced the data with our pipeline, yielding calibrated $R$ band spectra of
our candidates in the range 5800~\AA $\le$ $\lambda$ $\le$ 7400~\AA. Due to
the much higher sky background, however, the $R$ band spectra are considerably
noisier than the $B$ and $V$ band spectra.  Therefore, only emission lines of
brighter objects of our sample were unambiguously detected. Nevertheless, the
$R$ band spectra confirmed the previously adopted low redshift assignments of
8 candidates. No systematic search for further emission line objects was
performed among the $R$ band spectra.

\section{Spectroscopic follow-up}
\label{qnq_followup}

\subsection{Observations}
\label{qnq_obs}

Follow-up spectroscopy of 81 quasar candidates was obtained with the Focal
Reducer/Low Dispersion Spectrograph 2 \citep[FORS2,][]{appenzeller98} on ESO
VLT UT1/Antu in Visitor Mode on November 17--19, 2004 and in Service Mode
between April and July 2005. We restricted the follow-up campaign to candidates
with \textbf{a} high-confidence slitless WFI redshift $z\ge 1.7$. Quasar
candidates likely residing at lower redshifts (\ion{C}{iii}] or \ion{Mg}{ii}
seen in the slitless spectra) were not followed up. The field of PKS~0528$-$250
was not selected for follow-up and also the field of Q~1451$+$123 was not
included due to the lack of high-quality candidates. Longslit spectra with a
1\arcsec\ slit kept at the parallactic angle were taken either with the 300V
grism or the 600B grism, resulting in a spectral resolution of $\sim 10$~\AA\
FWHM and $\sim 4.5$~\AA\ FWHM, respectively. Each resolution element is
oversampled by a factor of $\sim 3$
(300V: $3.36$~\AA/pixel, 600B: 1.5~\AA/pixel). No order separation filter was
employed to maximise the UV coverage in the spectra, leading to possible order
overlap at $\lambda>6600$~\AA\ in the spectra taken with the 300V grism.
Exposure times were adjusted to yield $S/N\sim 20$ in the quasar continuum and
ranged between one and 40 minutes. The sky was clear to photometric during the
Visitor Mode run, but the seeing varied strongly during the nights, resulting
in slit losses for some of the 53 targeted objects. The remaining 28 prime
quasar candidates were observed in Service Mode under variable conditions. The
spectra were calibrated in wavelength against the FORS2 He/Ne/Ar/HgCd arc lamps
and flux-calibrated via spectrophotometric standard stars taken each night.
However, absolute spectrophotometry was not achieved due to the narrow slit and
the sometimes variable or poor sky transparency in the Service Mode
observations. Table~\ref{qnq_forsobservinglog}, available in the online edition
of the Journal, lists the quasars confirmed in the spectroscopic follow-up
observations.

\onllongtab{2}{
\onecolumn\centering\scriptsize
\begin{longtable}{lllllllllll}
\caption{\label{qnq_forsobservinglog}Observing log of the spectroscopic follow-up.
The quasars reported in Paper~I \& II have been included for completeness. The
columns list the quasar field, the name of the confirmed quasar, its right
ascension and declination, its redshift with estimated error, its $B$ magnitude
with $1\sigma$ error, night of observation, the grism used, the exposure time,
the seeing and the sky conditions during observation.}\\
\hline\hline\noalign{\smallskip}
Field		&QSO			&$\alpha$~(J2000) 		&$\delta$~(J2000)	&$z$	&$B$	&Night	&Grism	&$t_\mathrm{exp}$~[s]	&Seeing	&Transparency\\
\noalign{\smallskip}\hline\noalign{\smallskip}
\endfirsthead
\caption{continued.}\\
\hline\hline\noalign{\smallskip}
Field		&QSO			&$\alpha$~(J2000) 		&$\delta$~(J2000)	&$z$	&$B$	&Night	&Grism	&$t_\mathrm{exp}$~[s]	&Seeing	&Transparency\\
\noalign{\smallskip}\hline\noalign{\smallskip}
\endhead
\noalign{\smallskip}\hline
\endfoot
\noalign{\smallskip}\hline\noalign{\smallskip}\multicolumn{11}{l}{
$^\mathrm{a}$ Due to imperfect astrometry the position of QNQ~J03052$+$0000 given in Paper~I is incorrect by $\simeq 3\arcsec$. Here we give the coordinates measured in the SDSS.}
\endlastfoot
Q~0000$-$263	&\object{QNQ~J00025$-$2558}&00$^{\mathrm{h}}$02$^{\mathrm{m}}$34\fs83&$-$25\degr58\arcmin44\farcs0	&$0.887\pm0.002$&$20.26\pm0.13$	&18 Nov 2004	&300V	&300	&0\farcs9	&clear\\
                &\object{QNQ~J00040$-$2603}&00$^{\mathrm{h}}$04$^{\mathrm{m}}$05\fs33&$-$26\degr03\arcmin41\farcs9	&$2.002\pm0.003$&$20.54\pm0.16$	&18 Nov 2004	&600B	&1000	&1\farcs0	&clear\\
		&\object{QNQ~J00035$-$2610}&00$^{\mathrm{h}}$03$^{\mathrm{m}}$31\fs88&$-$26\degr10\arcmin54\farcs8	&$2.771\pm0.003$&$21.61\pm0.42$	&18 Nov 2004	&600B	&1800	&1\farcs0	&clear\\
		&\object{QNQ~J00028$-$2547}&00$^{\mathrm{h}}$02$^{\mathrm{m}}$53\fs86&$-$25\degr47\arcmin43\farcs1	&$2.812\pm0.004$&$19.84\pm0.10$ &18 Nov 2004	&600B	&400	&0\farcs8	&clear\\
		&\object{QNQ~J00035$-$2551}&00$^{\mathrm{h}}$03$^{\mathrm{m}}$33\fs84&$-$25\degr51\arcmin49\farcs8	&$2.875\pm0.004$&$20.99\pm0.25$ &18 Nov 2004	&600B	&1600	&1\farcs1	&clear\\
                &\object{QNQ~J00038$-$2617}&00$^{\mathrm{h}}$03$^{\mathrm{m}}$51\fs44&$-$26\degr17\arcmin37\farcs8	&$3.073\pm0.003$&$22.09\pm0.41$	&18 Nov 2004	&300V	&1200	&1\farcs0	&clear\\
Q~0002$-$422	&\object{QNQ~J00043$-$4151}&00$^{\mathrm{h}}$04$^{\mathrm{m}}$19\fs05&$-$41\degr51\arcmin10\farcs6	&$0.743\pm0.001$&$19.84\pm0.16$ &19 Nov 2004	&300V	&300	&1\farcs3	&photometric\\
                &\object{QNQ~J00041$-$4158}&00$^{\mathrm{h}}$04$^{\mathrm{m}}$09\fs01&$-$41\degr58\arcmin32\farcs4	&$1.720\pm0.004$&$20.36\pm0.23$ &19 Nov 2004	&300V	&400	&1\farcs3	&photometric\\
		&\object{QNQ~J00045$-$4201}&00$^{\mathrm{h}}$04$^{\mathrm{m}}$32\fs75&$-$42\degr01\arcmin33\farcs9	&$2.157\pm0.002$&$20.35\pm0.21$ &19 Nov 2004	&300V	&400	&1\farcs5	&photometric\\
Q~0055$-$269	&\object{QNQ~J00576$-$2626}&00$^{\mathrm{h}}$57$^{\mathrm{m}}$36\fs26&$-$26\degr26\arcmin57\farcs1	&$1.942\pm0.004$&$20.73\pm0.19$ &18 Nov 2004	&300V	&300	&1\farcs2	&clear\\   
                &\object{QNQ~J00582$-$2649}&00$^{\mathrm{h}}$58$^{\mathrm{m}}$13\fs94&$-$26\degr49\arcmin19\farcs2	&$2.572\pm0.003$&$22.02\pm0.44$ &18 Nov 2004	&300V	&1800	&0\farcs9	&clear\\
		&\object{QNQ~J00583$-$2626}&00$^{\mathrm{h}}$58$^{\mathrm{m}}$19\fs50&$-$26\degr26\arcmin12\farcs8	&$2.720\pm0.006$&$21.00\pm0.26$ &18 Nov 2004	&300V	&300	&1\farcs2	&clear\\
Q~0302$-$003	&\object{QNQ~J03052$-$0016}&03$^{\mathrm{h}}$05$^{\mathrm{m}}$15\fs62&$-$00\degr16\arcmin14\farcs4	&$2.290\pm0.002$&$20.05\pm0.06$ &17 Nov 2004	&300V	&300	&0\farcs7	&clear\\
                &\object{QNQ~J03052$+$0000}$\,^\mathrm{a}$
                                  &03$^{\mathrm{h}}$05$^{\mathrm{m}}$16\fs95&$+$00\degr00\arcmin43\farcs5	&$2.808\pm0.004$&$21.79\pm0.22$ &17 Nov 2004	&300V	&600	&0\farcs9	&clear\\
Q~0347$-$383	&\object{QNQ~J03494$-$3814}&03$^{\mathrm{h}}$49$^{\mathrm{m}}$24\fs35&$-$38\degr14\arcmin34\farcs2	&$1.471\pm0.002$&$19.60\pm0.04$ &17 Nov 2004	&300V	&60	&0\farcs8	&clear\\
                &\object{QNQ~J03500$-$3820}&03$^{\mathrm{h}}$50$^{\mathrm{m}}$04\fs26&$-$38\degr20\arcmin51\farcs2	&$1.819\pm0.002$&$20.03\pm0.05$ &17 Nov 2004	&300V	&120	&0\farcs6	&clear\\
		&\object{QNQ~J03490$-$3812}&03$^{\mathrm{h}}$49$^{\mathrm{m}}$03\fs60&$-$38\degr12\arcmin35\farcs8	&$1.945\pm0.003$&$21.66\pm0.19$ &17 Nov 2004	&300V	&600	&0\farcs9	&clear\\
		&\object{QNQ~J03496$-$3821}&03$^{\mathrm{h}}$49$^{\mathrm{m}}$39\fs31&$-$38\degr21\arcmin34\farcs1	&$2.351\pm0.003$&$19.43\pm0.03$ &17 Nov 2004	&600B	&300	&0\farcs7	&clear\\
		&\object{QNQ~J03496$-$3810}&03$^{\mathrm{h}}$49$^{\mathrm{m}}$36\fs28&$-$38\degr10\arcmin02\farcs1	&$2.433\pm0.003$&$20.87\pm0.09$ &17 Nov 2004	&600B	&1800	&0\farcs8	&clear\\
		&\object{QNQ~J03495$-$3806}&03$^{\mathrm{h}}$49$^{\mathrm{m}}$32\fs61&$-$38\degr06\arcmin45\farcs1	&$2.475\pm0.003$&$20.86\pm0.09$ &17 Nov 2004	&600B	&1800	&0\farcs7	&clear\\
		&\object{QNQ~J03508$-$3812}&03$^{\mathrm{h}}$50$^{\mathrm{m}}$50\fs70&$-$38\degr12\arcmin39\farcs0	&$2.705\pm0.003$&$20.86\pm0.06$ &17 Nov 2004	&300V	&360	&0\farcs6	&clear\\
		&\object{QNQ~J03503$-$3800}&03$^{\mathrm{h}}$50$^{\mathrm{m}}$20\fs01&$-$38\degr00\arcmin03\farcs6	&$2.734\pm0.003$&$20.67\pm0.07$ &17 Nov 2004	&600B	&900	&0\farcs6	&clear\\
		&\object{QNQ~J03490$-$3825}&03$^{\mathrm{h}}$49$^{\mathrm{m}}$02\fs58&$-$38\degr25\arcmin21\farcs5	&$2.777\pm0.003$&$20.52\pm0.06$ &17 Nov 2004	&600B	&1000	&0\farcs7	&clear\\
		&\object{QNQ~J03494$-$3826}&03$^{\mathrm{h}}$49$^{\mathrm{m}}$28\fs40&$-$38\degr26\arcmin11\farcs6	&$2.782\pm0.002$&$20.97\pm0.07$ &17 Nov 2004	&600B	&1800	&0\farcs6	&clear\\
CTQ~0247   	&\object{QNQ~J04061$-$4401}&04$^{\mathrm{h}}$06$^{\mathrm{m}}$10\fs38&$-$44\degr01\arcmin00\farcs0	&$2.410\pm0.020$&$19.35\pm0.05$ &19 Nov 2004	&600B	&900	&1\farcs0	&photometric\\             
		&\object{QNQ~J04075$-$4416}&04$^{\mathrm{h}}$07$^{\mathrm{m}}$35\fs34&$-$44\degr16\arcmin04\farcs1	&$3.034\pm0.003$&$21.75\pm0.19$ &19 Nov 2004	&600B	&2000	&1\farcs1	&photometric\\
		&\object{QNQ~J04084$-$4420}&04$^{\mathrm{h}}$08$^{\mathrm{m}}$29\fs02&$-$44\degr20\arcmin14\farcs4	&$3.080\pm0.004$&$22.88\pm0.41$ &19 Nov 2004	&300V	&1800	&1\farcs7	&photometric\\
Q~0420$-$388	&\object{QNQ~J04217$-$3847}&04$^{\mathrm{h}}$21$^{\mathrm{m}}$45\fs70&$-$38\degr47\arcmin44\farcs9	&$0.771\pm0.002$&$20.73\pm0.09$ &18 Nov 2004	&600B	&800	&0\farcs9	&clear\\
                &\object{QNQ~J04229$-$3831}&04$^{\mathrm{h}}$22$^{\mathrm{m}}$59\fs97&$-$38\degr31\arcmin37\farcs1	&$1.990\pm0.002$&$19.98\pm0.06$ &18 Nov 2004	&600B	&400	&0\farcs8	&clear\\
		&\object{QNQ~J04222$-$3829}&04$^{\mathrm{h}}$22$^{\mathrm{m}}$17\fs31&$-$38\degr29\arcmin33\farcs4	&$2.168\pm0.003$&$20.91\pm0.10$ &18 Nov 2004	&300V	&300	&0\farcs8	&clear\\
		&\object{QNQ~J04215$-$3857}&04$^{\mathrm{h}}$21$^{\mathrm{m}}$34\fs81&$-$38\degr57\arcmin03\farcs4	&$2.235\pm0.003$&$22.23\pm0.27$ &18 Nov 2004	&300V	&1400	&1\farcs4	&clear\\
                &\object{QNQ~J04215$-$3854}&04$^{\mathrm{h}}$21$^{\mathrm{m}}$32\fs16&$-$38\degr54\arcmin50\farcs6	&$2.270\pm0.010$&$19.49\pm0.04$ &18 Nov 2004	&600B	&200	&0\farcs7	&clear\\
		&\object{QNQ~J04212$-$3853}&04$^{\mathrm{h}}$21$^{\mathrm{m}}$12\fs49&$-$38\degr53\arcmin28\farcs5	&$2.723\pm0.003$&$20.08\pm0.06$ &18 Nov 2004	&600B	&1000	&0\farcs7	&clear\\
		&\object{QNQ~J04230$-$3853}&04$^{\mathrm{h}}$23$^{\mathrm{m}}$01\fs12&$-$38\degr53\arcmin14\farcs5	&$3.042\pm0.005$&$21.95\pm0.15$ &18 Nov 2004	&300V	&1800	&0\farcs7	&clear\\		
HE~0940$-$1050	&\object{QNQ~J09422$-$1117}&09$^{\mathrm{h}}$42$^{\mathrm{m}}$13\fs67&$-$11\degr17\arcmin37\farcs5	&$0.741\pm0.003$&$20.10\pm0.07$ &06 Apr 2005	&600B	&420	&0\farcs9	&thin cirrus\\
                &\object{QNQ~J09437$-$1109}&09$^{\mathrm{h}}$43$^{\mathrm{m}}$42\fs67&$-$11\degr09\arcmin47\farcs6	&$1.456\pm0.002$&$21.00\pm0.13$ &06 Apr 2005	&300V	&600	&0\farcs8	&thin cirrus\\
		&\object{QNQ~J09427$-$1108}&09$^{\mathrm{h}}$42$^{\mathrm{m}}$44\fs14&$-$11\degr08\arcmin00\farcs3	&$1.517\pm0.004$&$19.91\pm0.07$ &12 May 2005	&600B	&300	&1\farcs2	&photometric\\
		&\object{QNQ~J09430$-$1108}&09$^{\mathrm{h}}$43$^{\mathrm{m}}$00\fs27&$-$11\degr08\arcmin06\farcs3	&$1.730\pm0.003$&$21.03\pm0.12$ &10 May 2005	&300V	&1800	&0\farcs7	&clear\\
		&\object{QNQ~J09424$-$1047}&09$^{\mathrm{h}}$42$^{\mathrm{m}}$24\fs17&$-$10\degr47\arcmin58\farcs5	&$1.971\pm0.003$&$20.61\pm0.09$ &12 May 2005	&300V	&240	&0\farcs7	&photometric\\
		&\object{QNQ~J09437$-$1057}&09$^{\mathrm{h}}$43$^{\mathrm{m}}$45\fs15&$-$10\degr57\arcmin58\farcs8	&$2.023\pm0.004$&$20.64\pm0.10$ &09 Jun 2005	&300V	&240	&0\farcs6	&clear\\
		&\object{QNQ~J09435$-$1049}&09$^{\mathrm{h}}$43$^{\mathrm{m}}$30\fs04&$-$10\degr49\arcmin58\farcs9	&$2.216\pm0.003$&$20.79\pm0.12$ &06 Apr 2005	&600B	&720	&0\farcs9	&thin cirrus\\
		&\object{QNQ~J09425$-$1048}&09$^{\mathrm{h}}$42$^{\mathrm{m}}$30\fs58&$-$10\degr48\arcmin50\farcs8	&$2.325\pm0.005$&$19.80$	&02 May 2005	&600B	&720	&1\farcs2	&thick cirrus\\
		&\object{QNQ~J09434$-$1053}&09$^{\mathrm{h}}$43$^{\mathrm{m}}$24\fs21&$-$10\degr53\arcmin32\farcs9	&$2.760\pm0.003$&$21.16\pm0.18$ &12 May 2005	&300V	&1400	&1\farcs3	&photometric\\
		&\object{QNQ~J09427$-$1121}&09$^{\mathrm{h}}$42$^{\mathrm{m}}$44\fs42&$-$11\degr21\arcmin38\farcs9	&$2.963\pm0.003$&$20.99\pm0.10$ &19 Nov 2004	&600B	&1200	&1\farcs2	&photometric\\
                &&&&&&05 Apr 2005	&600B	&600	&0\farcs9	&thick cirrus\\
		&\object{QNQ~J09437$-$1052}&09$^{\mathrm{h}}$43$^{\mathrm{m}}$42\fs99&$-$10\degr52\arcmin31\farcs7	&$3.018\pm0.003$&$20.78\pm0.09$ &19 Nov 2004	&600B	&1200	&1\farcs3	&photometric\\
                &&&&&&06 Apr 2005	&600B	&1200	&0\farcs5	&thick cirrus\\
CTQ~0460	&\object{QNQ~J10399$-$2321}&10$^{\mathrm{h}}$39$^{\mathrm{m}}$58\fs77&$-$23\degr21\arcmin40\farcs3	&$2.216\pm0.004$&$20.73\pm0.16$ &07 May 2005	&600B	&900	&0\farcs6	&thin cirrus\\
                &\object{QNQ~J10388$-$2258}&10$^{\mathrm{h}}$38$^{\mathrm{m}}$50\fs12&$-$22\degr58\arcmin08\farcs9	&$2.326\pm0.003$&$19.63\pm0.07$ &28 Jun 2005	&600B	&420	&0\farcs7	&thin cirrus\\
                &\object{QNQ~J10385$-$2317}&10$^{\mathrm{h}}$38$^{\mathrm{m}}$31\fs54&$-$23\degr17\arcmin55\farcs3	&$3.099\pm0.004$&$21.28\pm0.11$ &09 Jun 2005	&600B	&1800	&0\farcs6	&clear\\
BR~1117$-$1329	&\object{QNQ~J11208$-$1345}&11$^{\mathrm{h}}$20$^{\mathrm{m}}$48\fs50&$-$13\degr45\arcmin35\farcs6	&$1.893\pm0.002$&$20.10\pm0.08$ &27 Jun 2005	&300V	&180	&0\farcs9	&thin cirrus\\
                &\object{QNQ~J11205$-$1343}&11$^{\mathrm{h}}$20$^{\mathrm{m}}$34\fs09&$-$13\degr43\arcmin28\farcs9	&$1.910\pm0.002$&$19.84\pm0.07$ &27 Jun 2005	&300V	&180	&0\farcs8	&thin cirrus\\
		&\object{QNQ~J11197$-$1340}&11$^{\mathrm{h}}$19$^{\mathrm{m}}$46\fs71&$-$13\degr40\arcmin47\farcs1	&$2.220\pm0.004$&$18.32\pm0.04$ &27 Jun 2005	&600B	&420	&0\farcs9	&thin cirrus\\
		&\object{QNQ~J11192$-$1334}&11$^{\mathrm{h}}$19$^{\mathrm{m}}$12\fs37&$-$13\degr34\arcmin05\farcs5	&$3.252\pm0.005$&$22.55\pm0.40$ &27 Jun 2005	&300V	&1600	&0\farcs8	&thin cirrus\\
BR~1202$-$0725	&\object{QNQ~J12059$-$0754}&12$^{\mathrm{h}}$05$^{\mathrm{m}}$55\fs33&$-$07\degr54\arcmin33\farcs7	&$0.773\pm0.001$&$20.40\pm0.20$ &06 May 2005	&600B	&900	&1\farcs5	&clear\\
                &\object{QNQ~J12062$-$0727}&12$^{\mathrm{h}}$06$^{\mathrm{m}}$14\fs24&$-$07\degr27\arcmin41\farcs1	&$1.478\pm0.003$&$20.46\pm0.08$ &29 Jun 2005	&300V	&180	&0\farcs9	&thin cirrus\\
                &\object{QNQ~J12061$-$0745}&12$^{\mathrm{h}}$06$^{\mathrm{m}}$08\fs91&$-$07\degr45\arcmin49\farcs8	&$1.730\pm0.002$&$19.15\pm0.06$ &06 May 2005	&600B	&600	&1\farcs5	&clear\\
Q~1209$+$093	&\object{QNQ~J12124$+$0851}&12$^{\mathrm{h}}$12$^{\mathrm{m}}$29\fs08&$+$08\degr51\arcmin58\farcs9	&$0.729\pm0.003$&$19.61\pm0.05$ &28 Jun 2005	&600B	&300	&0\farcs8	&thin cirrus\\
                &\object{QNQ~J12111$+$0906}&12$^{\mathrm{h}}$11$^{\mathrm{m}}$06\fs59&$+$09\degr06\arcmin43\farcs6	&$2.534\pm0.002$&$21.17\pm0.15$ &29 Jun 2005	&600B	&2400	&0\farcs7	&thin cirrus\\
PKS~2126$-$158	&\object{QNQ~J21294$-$1521}&21$^{\mathrm{h}}$29$^{\mathrm{m}}$29\fs18&$-$15\degr21\arcmin57\farcs9	&$0.580\pm0.001$&$18.38$ 	&04 May 2005	&600B	&180	&0\farcs8	&thin cirrus\\
                &\object{QNQ~J21291$-$1524A}&21$^{\mathrm{h}}$29$^{\mathrm{m}}$10\fs09&$-$15\degr24\arcmin45\farcs1	&$0.782\pm0.001$&$21.81\pm0.30$ &04 May 2005	&300V	&2400	&0\farcs9	&thin cirrus\\
		&\object{QNQ~J21297$-$1536}&21$^{\mathrm{h}}$29$^{\mathrm{m}}$46\fs75&$-$15\degr36\arcmin51\farcs9	&$1.509\pm0.003$&$20.50\pm0.13$ &06 May 2005	&300V	&180	&1\farcs3	&clear\\
                &\object{QNQ~J21286$-$1528}&21$^{\mathrm{h}}$28$^{\mathrm{m}}$40\fs43&$-$15\degr28\arcmin15\farcs7	&$1.925\pm0.005$&$19.79\pm0.10$ &06 May 2005	&300V	&120	&1\farcs6	&clear\\
		&\object{QNQ~J21291$-$1524B}&21$^{\mathrm{h}}$29$^{\mathrm{m}}$10\fs85&$-$15\degr24\arcmin23\farcs7	&$2.480\pm0.030$&$20.33\pm0.12$ &04 May 2005	&600B	&720	&1\farcs0	&thin cirrus\\
		&\object{QNQ~J21301$-$1533}&21$^{\mathrm{h}}$30$^{\mathrm{m}}$07\fs46&$-$15\degr33\arcmin20\farcs9	&$3.487\pm0.003$&$21.94\pm0.32$ &05 May 2005	&600B	&2200	&0\farcs9	&clear\\
Q~2139$-$4434	&\object{QNQ~J21434$-$4432}&21$^{\mathrm{h}}$43$^{\mathrm{m}}$25\fs26&$-$44\degr32\arcmin11\farcs0	&$2.709\pm0.004$&$19.33\pm0.06$ &05 May 2005	&600B	&540	&1\farcs1	&clear\\
HE~2243$-$6031	&\object{QNQ~J22454$-$6020}&22$^{\mathrm{h}}$45$^{\mathrm{m}}$27\fs22&$-$60\degr20\arcmin25\farcs2	&$1.984\pm0.015$&$20.94\pm0.29$ &19 Nov 2004	&300V	&300	&1\farcs6	&photometric\\
                &\object{QNQ~J22455$-$6015}&22$^{\mathrm{h}}$45$^{\mathrm{m}}$34\fs64&$-$60\degr15\arcmin45\farcs9	&$2.036\pm0.002$&$21.74\pm0.42$ &19 Nov 2004	&300V	&800	&1\farcs2	&photometric\\
		&\object{QNQ~J22460$-$6024}&22$^{\mathrm{h}}$46$^{\mathrm{m}}$01\fs23&$-$60\degr24\arcmin57\farcs5	&$2.041\pm0.003$&$21.25\pm0.33$ &19 Nov 2004	&300V	&1800	&1\farcs1	&photometric\\
		&\object{QNQ~J22454$-$6011}&22$^{\mathrm{h}}$45$^{\mathrm{m}}$29\fs42&$-$60\degr11\arcmin17\farcs1	&$2.324\pm0.003$&$19.36\pm0.06$ &17 Nov 2004	&600B	&300	&0\farcs9	&clear\\
		&\object{QNQ~J22463$-$6009}&22$^{\mathrm{h}}$46$^{\mathrm{m}}$18\fs47&$-$60\degr09\arcmin02\farcs5	&$2.329\pm0.003$&$19.71\pm0.09$ &17 Nov 2004	&600B	&500	&1\farcs0	&clear\\
		&\object{QNQ~J22484$-$6002}&22$^{\mathrm{h}}$48$^{\mathrm{m}}$29\fs20&$-$60\degr02\arcmin19\farcs4	&$3.586\pm0.002$&$20.97\pm0.20$ &17 Nov 2004	&600B	&600	&1\farcs0	&clear\\
		&&&&&&19 Nov 2004	&600B	&600	&1\farcs4	&photometric\\\pagebreak
HE~2347$-$4342	&\object{QNQ~J23510$-$4336}&23$^{\mathrm{h}}$51$^{\mathrm{m}}$05\fs50&$-$43\degr36\arcmin57\farcs2	&$0.720\pm0.002$&$20.74\pm0.27$ &19 Nov 2004	&300V	&1200	&1\farcs3	&photometric\\
                &\object{QNQ~J23507$-$4319}&23$^{\mathrm{h}}$50$^{\mathrm{m}}$44\fs97&$-$43\degr19\arcmin26\farcs0	&$0.850\pm0.003$&$19.90\pm0.07$ &17 Nov 2004	&600B	&360	&0\farcs7	&clear\\
		&\object{QNQ~J23507$-$4326}&23$^{\mathrm{h}}$50$^{\mathrm{m}}$45\fs39&$-$43\degr26\arcmin37\farcs0	&$1.635\pm0.003$&$21.05\pm0.14$ &17 Nov 2004	&300V	&200	&1\farcs0	&clear\\
		&\object{QNQ~J23509$-$4330}&23$^{\mathrm{h}}$50$^{\mathrm{m}}$54\fs80&$-$43\degr30\arcmin42\farcs2	&$1.762\pm0.004$&$18.23\pm0.03$ &17 Nov 2004	&600B	&300	&0\farcs7	&clear\\
		&\object{QNQ~J23502$-$4334}&23$^{\mathrm{h}}$50$^{\mathrm{m}}$16\fs18&$-$43\degr34\arcmin14\farcs7	&$1.763\pm0.003$&$18.95\pm0.04$ &17 Nov 2004	&300V	&60	&0\farcs7	&clear\\
		&\object{QNQ~J23503$-$4328}&23$^{\mathrm{h}}$50$^{\mathrm{m}}$21\fs55&$-$43\degr28\arcmin43\farcs7	&$2.282\pm0.003$&$20.66\pm0.11$ &17 Nov 2004	&300V	&400	&0\farcs7	&clear\\
		&\object{QNQ~J23495$-$4338}&23$^{\mathrm{h}}$49$^{\mathrm{m}}$34\fs53&$-$43\degr38\arcmin08\farcs7	&$2.690\pm0.006$&$20.21\pm0.17$ &19 Nov 2004	&300V	&360	&1\farcs2	&photometric\\
		&\object{QNQ~J23511$-$4319}&23$^{\mathrm{h}}$51$^{\mathrm{m}}$09\fs44&$-$43\degr19\arcmin41\farcs6	&$3.020\pm0.004$&$21.00\pm0.14$ &17 Nov 2004	&600B	&1000	&1\farcs1	&clear\\
                &\object{QNQ~J23514$-$4339}&23$^{\mathrm{h}}$51$^{\mathrm{m}}$25\fs54&$-$43\degr39\arcmin02\farcs9	&$3.240\pm0.004$&$21.57\pm0.29$ &17 Nov 2004	&300V	&1400	&1\farcs2	&clear\\
		&\object{QNQ~J23503$-$4317}&23$^{\mathrm{h}}$50$^{\mathrm{m}}$21\fs94&$-$43\degr17\arcmin30\farcs0	&$3.542\pm0.005$&$21.94\pm0.62$ &19 Nov 2004	&300V	&1800	&1\farcs2	&photometric\\
		&&&&&	&19 Nov 2004	&600B	&1800	&1\farcs2	&photometric\\
\end{longtable}
}

\subsection{Data reduction}

The VLT spectra were reduced with standard IRAF\footnote{IRAF is distributed
  by the National Optical Astronomy Observatories, which are operated by the
  Association of Universities for Research in Astronomy, Inc., under
  cooperative agreement with the National Science Foundation.}  
tasks. The bias value was taken from the overscan regions. Dome flatfields
were used to correct pixel sensitivity variations. After sky subtraction by
fitting both spatial regions close to the target by 2nd-order Chebychev
polynomials, the spectra were extracted with the optimal extraction algorithm
by \citet{horne86}.  The root-mean-square residuals of the wavelength
calibration with low-order Chebychev polynomials were $\la 0.05$~\AA\ and
$\sim 0.2$~\AA\ for the 600B and the 300V grism, respectively. The spectra were
corrected for atmospheric extinction assuming average conditions. 

\begin{figure}
\resizebox{\hsize}{!}{\includegraphics{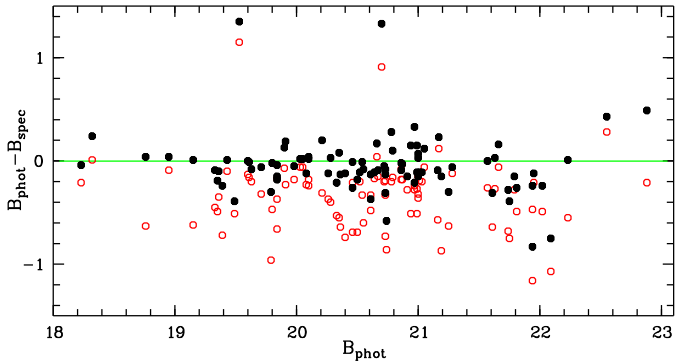}}
\caption{\label{qnq_photcorr} Blue magnitudes of the quasar candidates
  observed with the VLT. The abscissa shows the photometric $B$ magnitudes
  obtained from the aquisition images, the ordinate the difference to
  the magnitudes integrated from the slit spectra. Full (open) circles
  correspond to measurements with (without) correction for slit losses.
  The line marks a perfect 1:1 relation between $B_\mathrm{phot}$ and
  $B_\mathrm{spec}$.}
\end{figure}

In order to approximately correct the spectra for slit losses we carried out
aperture photometry on the FORS2 aquisition images taken in the $B$ filter,
which were calibrated against photometric standard stars \citep{landolt92}
taken each night.
Table~\ref{qnq_forsobservinglog} provides the airmass-corrected Johnson $B$
magnitudes and their calculated standard deviations. Apparent magnitudes were
also derived by integrating the flux-calibrated slit spectra. Slit losses were
estimated by calculating the expected loss over the used 1\arcsec\ slit for a
Gaussian PSF with a FWHM equal to the seeing during the observations that we
measured on the aquisition images.  Figure~\ref{qnq_photcorr} presents the
correlation of the integrated and photometric magnitudes. The integrated
magnitudes are consistent with the photometric ones after correcting for slit
losses. 
The two outliers are the two quasars QNQ~J09425$-$1048 and QNQ~J21294$-$1521
that were observed in strongly variable clearly non-photometric conditions.

\section{Results}
\label{qnq_results}

\subsection{Redshift determination}
\label{qnq_redshifts}

Redshifts of all candidates observed with FORS2 were determined by taking into
account all detectable emission lines. Peak positions of the lines were
measured by eye and errors were estimated by considering the $S/N$ of the
lines, blending with other emission lines and the presence of absorption
lines, sky residuals or line asymmetries. We confirmed that in many objects
high-ionisation lines (Ly$\alpha$, \ion{N}{v}, \ion{Si}{iv}+\ion{O}{iv}],
\ion{C}{iv}) are blueshifted with respect to low-ionisation lines
\citep{gaskell82,tytler92,mcintosh99,richards02}. However, the low-ionisation
lines \ion{O}{i}+\ion{Si}{ii} and \ion{C}{ii} were often weak and noisy,
resulting in larger individual redshift errors. \ion{Mg}{ii} could be measured
only for quasars at $z\la2.30$ ($z\la 1.25$) taken with the 300V (600B)
grism. Therefore, we estimated the systemic redshift of each quasar by
weighting the measurements of individual lines, giving a lower weight to
high-ionisation lines or discarding them completely in case of large
blueshifts. The redshift uncertainty of each quasar was estimated from the
redshift scatter between the remaining lines and their centroiding errors. The
adopted quasar redshifts and their estimated errors are listed in
Table~\ref{qnq_forsobservinglog}.

\begin{figure}
\resizebox{\hsize}{!}{\includegraphics{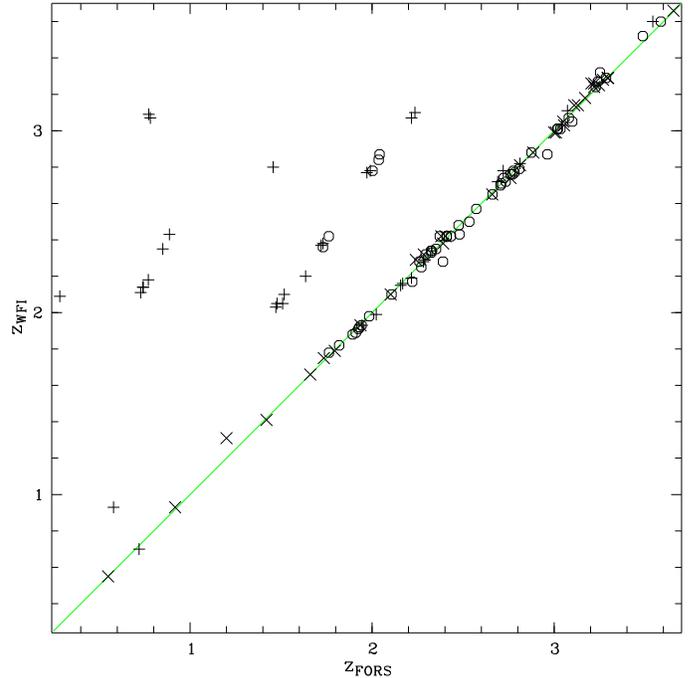}}
\caption{\label{qnq_wfiqnqcorr} Comparison of the redshifts obtained with FORS2
and WFI. Circles ($+$ signs) denote quasar candidates with almost secure
(uncertain) WFI redshifts. Previously known quasars in the fields have been
added with their published redshift in comparison to the redshift measured in
the slitless spectra ($\times$ signs). The line marks a 1:1 relation.}
\end{figure}

\subsection{Newly confirmed quasars}
\label{qnq_zcomparison}

\begin{figure*}
\centering
\includegraphics[width=\textwidth]{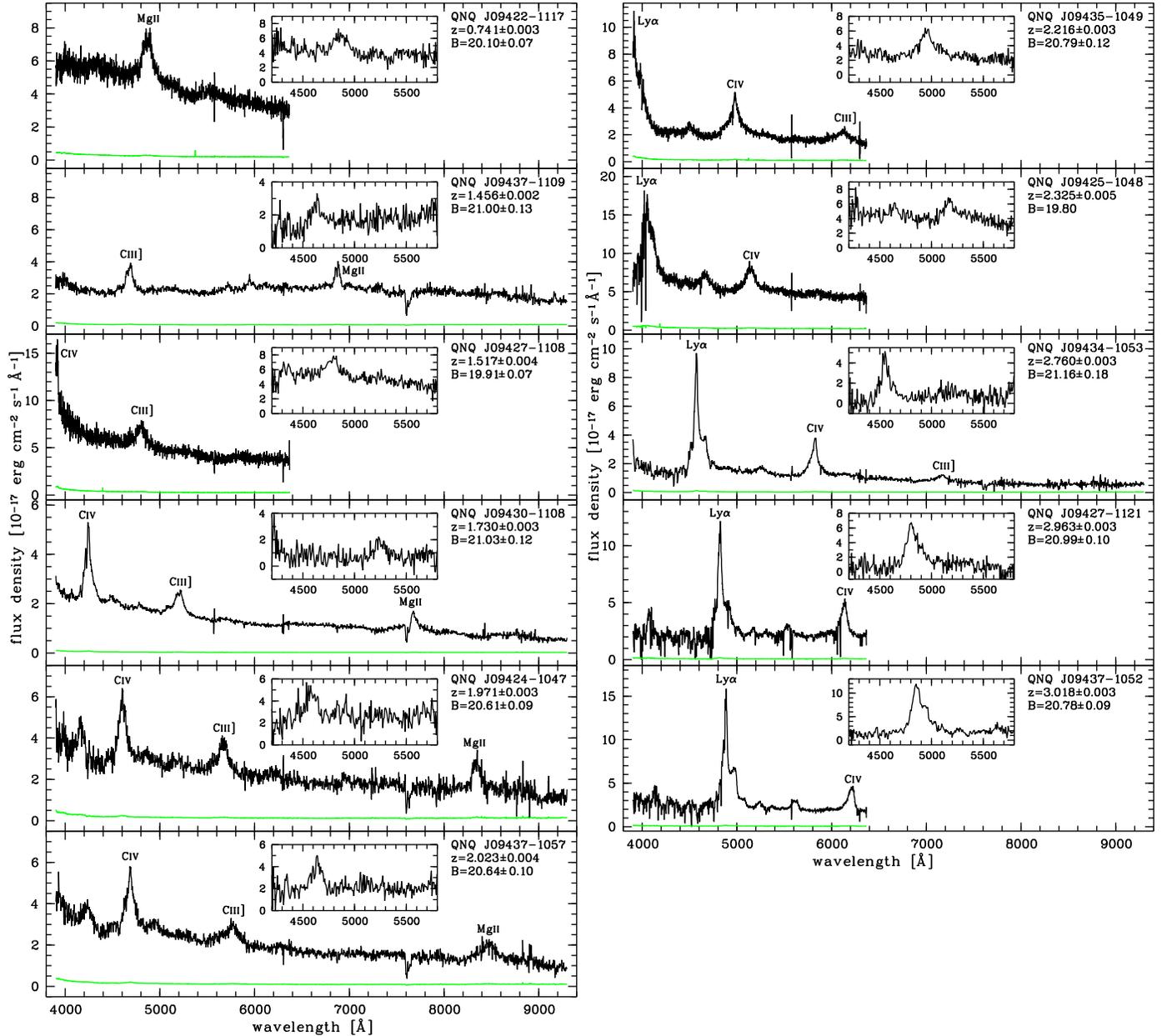}
\caption{\label{qnq_he0940forsspec} VLT/FORS2 spectra of the 11 discovered
  quasars in the vicinity of HE~0940$-$1050. The spectra are shown in black
  together with their $1\sigma$ noise arrays (green/grey). The spectra have
  been scaled to yield the measured photometric $B$ magnitudes. Major
  emission lines are labelled. The small inserts show the corresponding
  slitless WFI spectra in the same units.}
\end{figure*}

With the follow-up spectroscopy we confirmed that 80 of our 81 candidates are
broad-line AGN in the redshift range $0.580\le z\le 3.586$. Only one of the
candidates turned out to be a low-redshift galaxy with narrow emission lines
(object J03490$-$3820,
$\alpha$(J2000)=03$^{\mathrm{h}}$49$^{\mathrm{m}}$00\fs56,
$\delta$(J2000)$=-$38\degr20\arcmin31\farcs9, $z=0.2848$).  12 of the
confirmed quasars, located in the fields around Q~0302$-$003 and
HE~2347$-$4342, were already presented and discussed in Papers~I and II. In
Fig.~\ref{qnq_he0940forsspec} we show the spectra for the quasars in one
additional, particularly rich field. 11 quasars were found in the vicinity of
HE~0940$-$1050. Appendix~A, available in the online edition
of the Journal, lists the spectra of the remaining 57 newly discovered quasars.

In Fig.~\ref{qnq_wfiqnqcorr} we compare the redshifts derived from the FORS2
spectra and the slitless WFI spectra, respectively. The slit spectra confirmed
the redshifts determined from the slitless WFI spectra for the majority of
candidates (54/81). The WFI redshifts of the other 27 objects in the follow-up
had been overestimated (note that we had systematically assigned the highest
plausible redshifts). 64 of the 80 confirmed quasars are above our low-redshift
cutoff $z= 1.7$ set for inclusion in the follow-up.

The scatter between slit and slitless spectroscopic redshifts for the 54 quasars
correctly placed in $z$ is $\sigma_z\simeq 0.03$. The main cause for this
scatter is optical distortions of the grism causing the astrometric
transformation to produce residual variations of the wavelength zeropoint across
the WFI field of view.  Of the 27 candidates whose redshifts were overpredicted,
only 5 had been assigned a high confidence redshift. This suggests that the
majority of remaining candidates with high redshift confidence likely reside at
those redshifts. After completion of the follow-up observations we compared the
WFI spectra of quasars whose redshifts had been overestimated to the WFI spectra
of the remaining candidates and adjusted the redshift estimates where applicable.

\begin{figure}
\resizebox{\hsize}{!}{\includegraphics{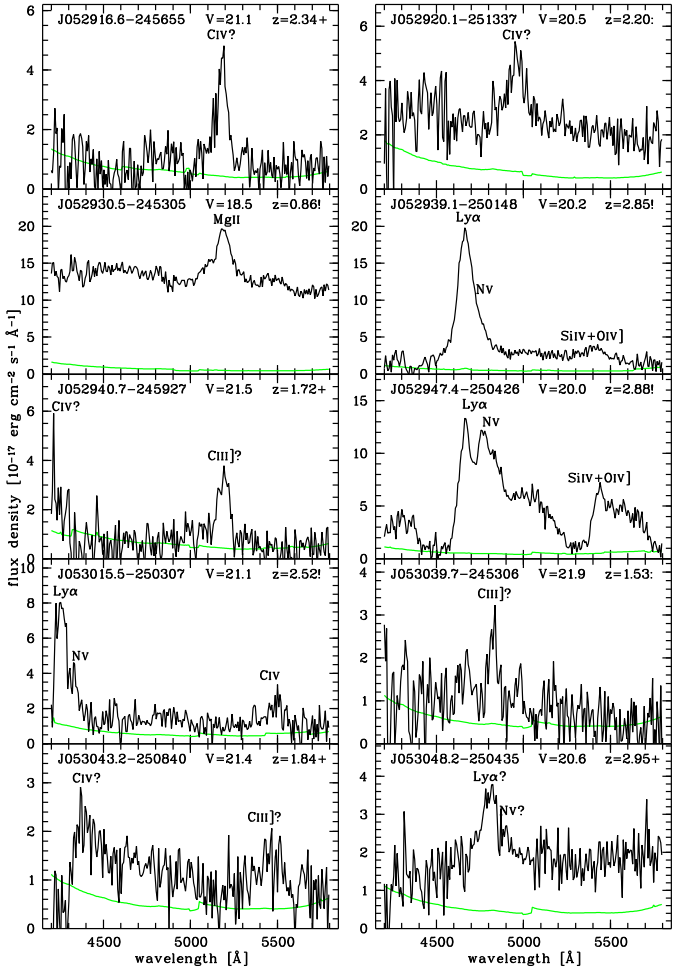}}
\caption{\label{qnq_pks0528cand} Slitless spectra (black) and corresponding
$1\sigma$ noise arrays (green/grey) of 10 quasar candidates in the field of
PKS~0528$-$250 with indicated coordinates, approximate $V$ magnitudes and
estimated redshifts with confidence level (!=secure, +=plausible, :=uncertain).
Major emission lines are labelled.}
\end{figure}

\subsection{Quasar candidates not included in the follow-up}

With 169 candidates in our sample and 81 of them followed up by VLT slit
spectroscopy, 88 further possible quasars remain unconfirmed. For the benefit of
potential future users we list these in  Appendix~B, available
in the online edition of the Journal. We provide the slitless spectra of these
88 quasar candidates and their celestial coordinates, magnitudes, WFI redshifts,
and our assessment of redshift confidence. For 23 of these we judge the WFI
redshifts to be most likely correct. For 28 candidates we call our redshift
estimate plausible; for 37 they are uncertain and could be identified with any
of the major emission lines, although we hold it unlikely that many Ly$\alpha$
quasars will be hidden among them. As an example we show in
Fig.~\ref{qnq_pks0528cand} the slitless spectra of 10 quasar candidates in the
field of PKS~0528$-$250 that was completely left out of the spectroscopic
follow-up. Three of these candidates have a secure redshift $z>2.5$. The
remaining candidates on the other fields are either faint with noisy WFI spectra
or they likely reside at redshifts $z<1.7$.

\begin{figure}
\centering
\includegraphics[angle=270,scale=1]{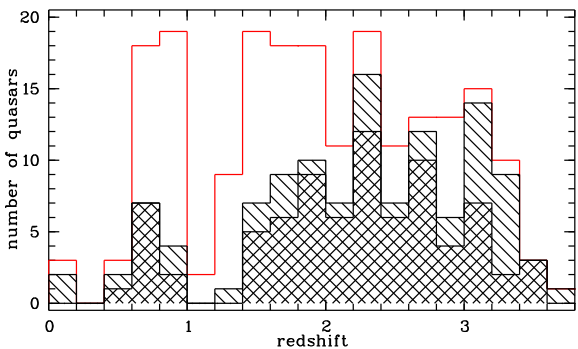}
\caption{\label{qnq_zhist} Redshift histograms of confirmed previously unknown
  quasars (cross-hashed), rediscovered quasars (hashed) and remaining quasar
  candidates (open).}
\end{figure}

\section{Discussion}
\label{qnq_discussion}

We summarise the overall redshift distribution of the quasar sample in
Fig.~\ref{qnq_zhist}. The open histogram shows the redshift distribution of all
detected known quasars, newly discovered quasars and remaining candidates
assuming that the inferred WFI redshifts of the candidates are correct. In
total, 205 quasars and candidates are shown. The gap at $1 \la z \la 1.3$
expectedly arises from the lack of visible quasar emission lines in the
observed spectral range (cf.\ Fig.~\ref{qnq_lineshift}). The very small number
of low-redshift ($z\la 0.6$) AGN is also to this effect. For $z\ga 1.4$, no
clearly preferred redshifts are discernible from this histogram. In
Sect.~\ref{qnq_completeness} we present an attempt to quantify our redshift-
and magnitude selection function. First we discuss our results in terms of the
main rationale driving this survey, the construction of new apparent groups of
high-redshift quasars.

\subsection{New quasar groups}

The main goal of this survey was to reveal new groups of quasars for further
studies of the three-dimensional distribution of the intergalactic medium. In
the follow-up observations we concentrated on likely high-redshift ($z>1.7$)
quasars, with some success: 80\% of the newly established quasars match that
condition. In addition to the central quasars, we recovered 85\% of all
detectable previously known quasars in these fields (22/26).

\begin{figure*}
\centering
\includegraphics[width=18.0cm]{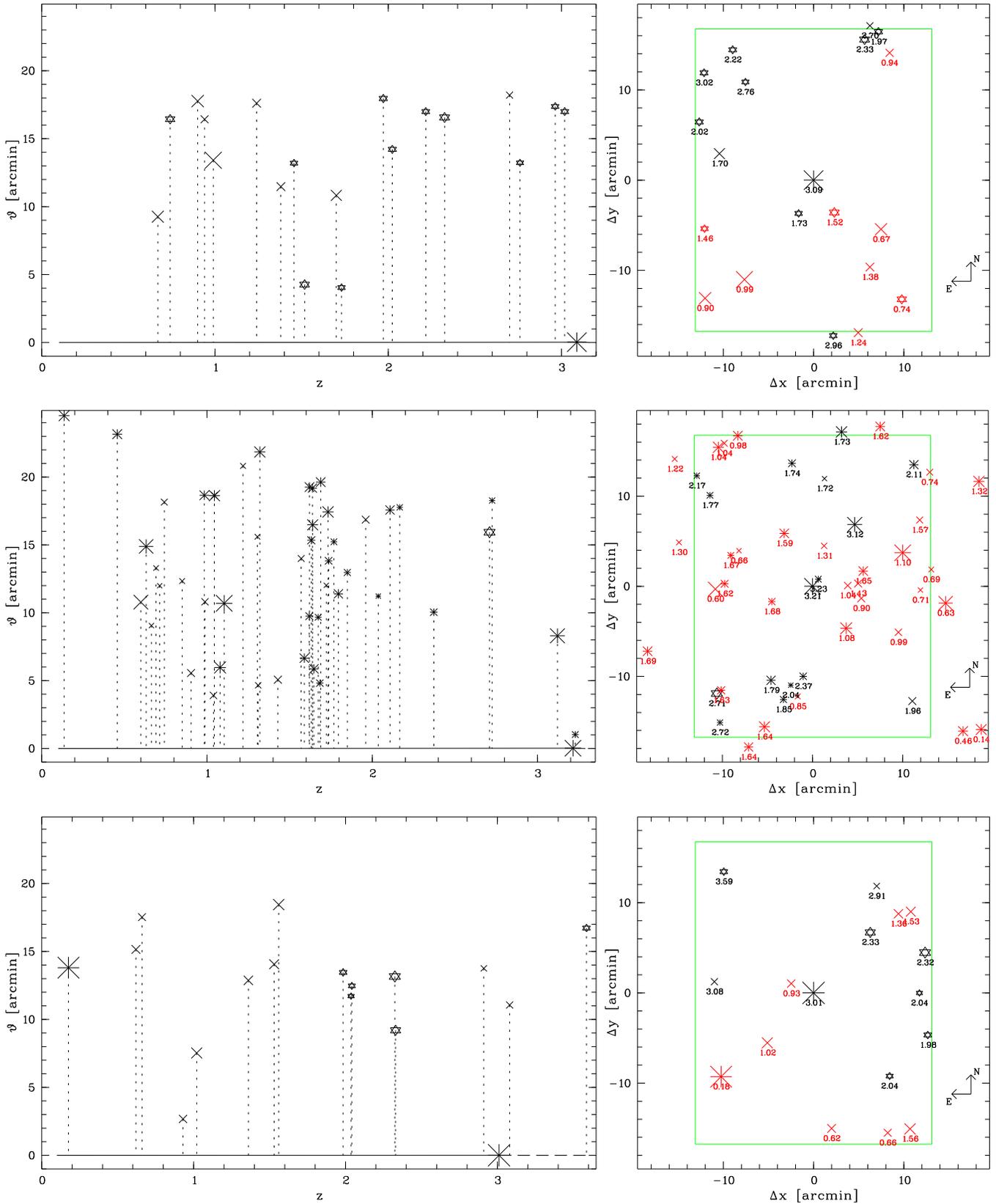}
\caption{\label{combfieldplot} Distributions of quasars in three of the survey
fields (\textit{upper panels:} HE~0940$-$1050, \textit{middle panels:}
Q~2139$-$4434, \textit{lower panels:} HE~2243$-$6031).
\textit{Left panels:} Angular separation distribution of quasars and quasar
candidates with respect to the central quasar as a function of redshift.
Asterisks and star symbols mark known quasars and newly discovered confirmed
quasars, respectively. Crosses show remaining WFI quasar candidates. Symbol size
indicates apparent optical magnitude.
\textit{Right panels:} Quasar distribution on the sky centred on the targeted
known quasar in the field with indicated redshifts. Objects at $z\ge 1.7$
($z<1.7$) are shown in black (red/grey). The rectangle marks the nominal
contiguous slitless WFI field of view ($26\farcm 2\times 33\farcm 5$) without
rotating the instrument, for simplicity centred on the central quasar. New
quasars located slightly outside the nominal field of view are due to field
rotation and small offsets of the actual field centre.}
\end{figure*}

Including the remaining candidates, the number of $z\ge 1.7$ quasars with
$V\la 21$ per $26\farcm2\times 33\farcm5$ field varies substantially, between 2
(near CTQ~0247, Q~1209+093 and Q~1451$+$123) and 11 (near Q~0347$-$383). Some
of our fields are thus almost devoid of high-redshift quasars except the central
object, but other fields we found to be quite richly populated. These fields may
in the future become interesting targets for multiple line of sight absorption
spectroscopy. A list of currently known $z\ge 1.7$ quasars in our survey fields
is provided in Table~C.1, available in the online edition of the
Journal. We briefly discuss some of the most prominent cases:

\textsl{HE~0940$-$1050 ($z=3.088$, $V=16.4$)}:
The high-quality VLT/UVES and Keck/HIRES spectra of HE~0940$-$1050 have been
heavily used in recent quasar absorption line studies, for example to constrain
the evolution of the fine structure constant \citep{murphy03,chand04} and to
obtain precise measurements of the flux distribution in the Ly$\alpha$ forest
\citep{becker07,kim07b}. So far, no other quasars or AGN have been known within
a radius $<30\arcmin$ around HE~0940$-$1050. We confirmed 11 out of 19 quasar
candidates. In the upper panels of Fig.~\ref{combfieldplot} we plot the angular
distribution of quasars and quasar candidates with respect to HE~0940$-$1050
together with the distribution of the quasars on the sky. Six confirmed
foreground quasars of HE~0940$-$1050 are located at $z>2$. The newly established
quasars in this field form two close subgroups on the sky with separations of
$1\farcm8<\vartheta<8\farcm8$. Thus, this field is well suited to study
transverse correlations in the IGM on small scales towards the newly discovered
quasars in addition to the larger scales given by their separations to the
central quasar. 

\textsl{Q~0302$-$003 ($z=3.285$, $V=17.7$)} and
\textsl{HE~2347$-$4342 ($z=2.885$, $V=16.4$)}:
These two quasars are among the very few for which intergalactic \ion{He}{ii}
absorption has been detected by UV spectroscopy
\citep[e.g.][]{jakobsen94,reimers97,heap00,smette02,zheng04}. The association
between the hardness of the intergalactic radiation field
(inferred from \ion{H}{i} and \ion{He}{ii} Ly$\alpha$ spectroscopy) and the
presence of foreground quasars near the line of sight has been explored in
Papers~I and II, revealing for the first time the `transverse proximity effect
in spectral hardness'. In Paper~II we also noted the remarkable apparent group
of 6 quasars with $z>2.2$ that is less than $17\arcmin$ from the central object,
3 of which are even behind HE~2347$-$4342.

\textsl{Q~0000$-$263 ($z=4.125$, $R=17.5$)}:
A $3\farcm 69\times 5\farcm 13$ region around this luminous high-redshift quasar
was included in the Lyman break galaxy survey by \citet{steidel03}, yielding
also 3 $g>24$ AGN at $z>3$. While these are obviously too faint to be
rediscovered, our survey revealed 4 quasars at $2.77<z<3.07$ at larger
separations to the central quasar.

\textsl{Q~0347$-$383 ($z=3.220$, $V=17.7$)}:
Among the 10 confirmed quasars around Q~0347$-$383 there is a dense group of 4
quasars at $2.70<z<2.78$ with magnitudes $20.5<B<21.0$. Thus, these quasars are
well suited for medium-resolution spectroscopy of the Ly$\alpha$ forest on
closely spaced sightlines. Two further $B\simeq 20.9$ quasars at $z=2.433$ and
$z=2.475$ are located within a radius of 5\arcmin\ around Q~0347$-$383.

\textsl{Q~0420$-$388 ($z=3.120$, $V=16.9$)}:
Taken in the ESO Large Programme
``The Cosmic Evolution of the Intergalactic Medium'' \citep{bergeron04}, the
UVES spectrum of Q~0420$-$388 is one of the highest-quality quasar spectra taken
so far (S/N$\simeq 100$ per $0.05$~\AA\ pixel).
Five out of seven newly established foreground quasars near Q~0420$-$388 are
located at $z>2$. The known quasar Q~0420$-$3850 \citep[$z=2.410$,][]{hewitt93}
separated from Q~0420$-$388 by $\vartheta=2\farcm 1$ was readily rediscovered in
our survey. In total, this field now contains seven $z>2$ quasars and two other
candidates with estimated high redshifts.

\textsl{Q~2139$-$4434 ($z=3.214$, $V=17.70$)}:
The sky region near this quasar (middle panels of Fig.~\ref{combfieldplot}) has
been intensely studied in previous surveys. \citet{morris91} discovered both
the central quasar Q~2139$-$4434 and another nearby bright high-redshift quasar
(Q~2138$-$4427, $z=3.120$, $\vartheta=8\farcm3$). Deeper surveys revealed
several fainter quasars \citep{veron95,hawkins00}. \citet{dodorico02} used
high-resolution spectra of the triplet formed by Q~2139$-$4434, its nearby
companion Q~2139$-$4433 ($z=3.228$, $\vartheta=1\arcmin$) and Q~2138$-$4427 in
order to reveal strong clustering of high-column density systems in the IGM in
the plane of the sky. More recently, \citet{francis04} presented follow-up
spectroscopy of colour-selected quasar and Ly$\alpha$ emitter candidates near
Q~2139$-$4434, finding an overdensity of $z\simeq 1.65$ quasars and a
large-scale filament of Ly$\alpha$ emitting galaxies at $z=2.38$. Besides
recovering the majority of the known quasars in this field, we identified a
further quite bright high-redshift quasar
(QNQ~J21434$-$4432, $B=19.3$, $z=2.709$). \citet{francis04} discovered two
fainter quasars at approximately the same redshift ($z=2.725$). One of them is
located just 3\farcm2 south of QNQ~J21434$-$4432. In total, there are now 14
confirmed $z>1.7$ quasars within a radius $<20\arcmin$ around Q~2139$-$4434.

\textsl{HE~2243$-$6031 ($z=3.010$, $V=16.4$)}:
The quasar distribution around HE~2243$-$6031 is shown in the lowest panels of
Fig.~\ref{combfieldplot}. This group is of particular interest for further
studies of quasar-absorber clustering. Near HE~2243$-$6031 our survey revealed
two bright quasars
(QNQ~J22463$-$6009 and QNQ~J22454$-$6011 from Table~\ref{qnq_forsobservinglog})
that coincide in redshift with the $z=2.33$ damped Ly$\alpha$ absorber towards
HE~2243$-$6031 discussed by \citet{lopez02}. The foreground quasars are
separated from the central line of sight by $9\farcm 2$ and $13\farcm 2$,
corresponding to transverse comoving distances of $15.1$ and $21.6$~Mpc in a
flat universe with density parameters
$\left(\Omega_\mathrm{m},\Omega_\Lambda\right)=\left(0.3,0.7\right)$ and a Hubble
constant $H_0=70$~km~s$^{-1}$~Mpc~$^{-1}$, respectively. Besides, this field
contains another probably associated quasar pair at $z\simeq 2.04$. Moreover,
the background quasar QNQ~J22484$-$6002 ($z=3.586$, $V=19.3$) is well suited for
high-resolution follow-up spectroscopy. This underlines the potential of our
present quasar sample and the need for further investigations.

For the remaining fields plots corresponding to Fig.~\ref{combfieldplot} are
presented in Appendix~C, available in the online edition of
the Journal.

\subsection{Quantification of selection effects}
\label{qnq_completeness}

We have shown above that quasars displaying strong emission lines can be
selected from the slitless spectra with very high efficiency; only 1 out of 81
candidates followed up with the VLT turned out not to be a quasar. Although in
all parts of the survey we optimised the procedure towards efficiency, also at
the expense of completeness, nevertheless it was interesting to see how our
emission-line based candidate selection performed with respect to the quasar
population displaying broad emission lines. We address this issue in two ways:
First we present an estimate of the survey selection function from Monte Carlo
simulations. Then we use this selection function to determine quasar surface
densities from our sample and compare these numbers with the results of other
surveys.

\subsubsection{Survey selection function}

We define the survey selection function as the recovered fraction of the
broad-line quasar population matching a given broad band flux limit. The main
variables governing the detectability of quasars in a survey such as ours are
(i) the continuum magnitudes and (ii) the equivalent widths of the detectable
emission lines \citep{schmidt86a,gratton87}. Quasar surveys based on emission
lines tend to preferentially select strong-lined quasars, even more so near
their survey limits. In order to quantify this selection effect, we first
generated several template quasar spectra via a Monte Carlo routine.
The quasar emission lines of Ly$\alpha$, \ion{Si}{iv}+\ion{O}{iv}], \ion{C}{iv},
\ion{C}{iii}] and \ion{Mg}{ii} were modelled as Gaussian profiles with mean
equivalent widths and line dispersions from the quasar template spectrum by
\citet{vandenberk01} superposed on a power law continuum
$f_\nu\propto\nu^{-\alpha}$ with a mean spectral index $\alpha=0.5$. We
incorporated variations in the model parameters by assuming Gaussian
distributions with a standard deviation of 20\% around the mean. We then
generated 200 random spectral templates in the quasar rest frame. The neglect of
non-Gaussian line shapes and intrinsic or intervening absorption does not have a
big effect on our low-resolution spectra.

Mock slitless WFI quasar spectra were created by shifting the templates to the
desired redshift, degrading them to the WFI resolution followed by adding
Gaussian noise that varied with wavelength according to the overall throughput
of the spectrograph (Fig.~\ref{qnq_filtercurves}).  The S/N in the simulated
spectra, related to the $V$ magnitude by eq.~(\ref{qnqeq_vsn}), was normalised
in the quasar continuum near the maximum throughput at 5400~\AA. We then
performed Monte Carlo simulations of the detection rates of quasars as a
function of redshift ($0.6\le z\le 3.7$) and continuum S/N values ($0.5\le
S/N\le 50$). At each considered redshift and S/N, WFI spectra of the 200
quasar templates were generated and subjected to our automated selection
routines (Sect.~\ref{qnq_selectcrit}). Each quasar template was simulated 100
times in order to quantify the impact of noise on the detection rate per object.
We also performed the subsequent visual screening done for the real candidates
on several hundred simulated spectra to investigate visual selection effects,
finding that $\ga 90$\% of the automatically selected candidates survived the
selection by eye.

\begin{figure}
\resizebox{\hsize}{!}{\includegraphics[angle=270]{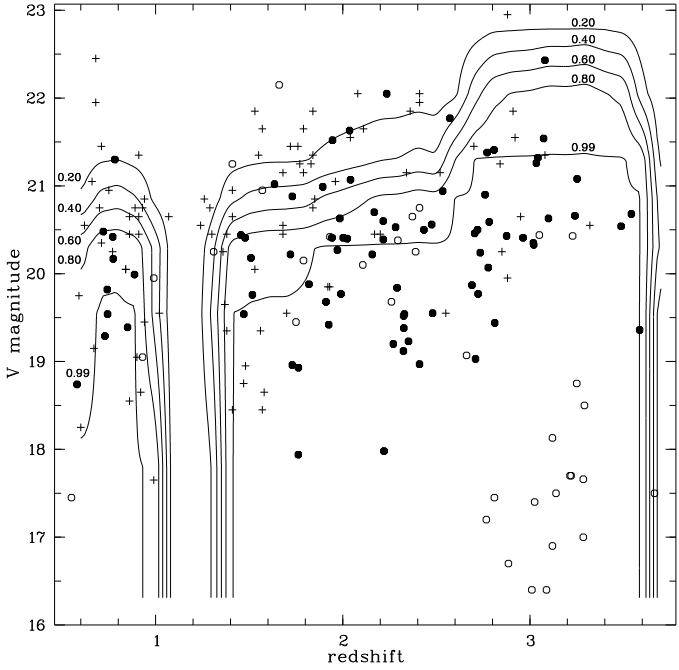}}
\caption{\label{qnq_completeness4} Survey selection function as a function of
redshift and apparent $V$ magnitude. Contours correspond to median simulated
detection probabilities $p\in\left\lbrace0.2,0.4,0.6,0.8,0.99\right\rbrace$.
Full and open circles mark newly discovered confirmed quasars and known quasars,
respectively. Crosses denote remaining candidates.}
\end{figure}

Figure~\ref{qnq_completeness4} presents the resulting selection function,
displayed as a contour map of the median detection probability as a function of
redshift and $V$ magnitude. Overplotted are the quasars and quasar candidates
from our survey. The selection function obviously varies with the visibility of
different emission lines in the spectrum. Due to its high equivalent width
Ly$\alpha$ emission is detectable for essentially 100\% of quasars brighter
than $V\simeq 21$, and still $\sim 80$\% of the simulated spectra are recovered
at $V\sim 22$ (continuum S/N of $\sim 1$). At redshifts below $\sim 2.5$ the
selection function obtains lower values because of the lower equivalent widths 
of \ion{C}{iv}, \ion{C}{iii}], and \ion{Mg}{ii}, and we have already commented
upon the gap at  $1.05\la z\la 1.30$ where no emission line is  visible in the
range of our spectra. The selected quasars and quasar candidates fill the full
range of the selection function spanned by the simulated redshifts and
magnitudes (S/N). The known quasars at the field centres are clearly separated
from the other quasars and candidates in magnitude and redshift. Note, however,
that the survey selection function in Fig.~\ref{qnq_completeness4} assumes that
all quasar spectra were extracted from the slitless data. Due to photometric
incompleteness (Sect.~\ref{qnq_surveydepth}), the detection probability at
$V\ga 21$ is lower, as we will describe in the following.

\subsubsection{Surface densities}

With the survey selection function from above we were able to compare our number
counts of quasars with those of other surveys. We consider two surveys reaching
similar or greater depths. The only comparably deep \emph{slitless} survey is
the Palomar Scan Grism Survey \citep[PSGS,][]{schneider99} which yielded 39
quasars on 1.10~deg$^2$ at $R\la 22$. Disregarding the known bright quasars at
the field centres we detected 102 quasars (known or newly confirmed) and 88
highly promising candidates on 4.39~deg$^2$. Assuming that the candidates are
quasars, the total surface density for our survey is $\simeq 20$\% higher. The
wavelength ranges and therefore the redshift ranges are not exactly equal
between the two surveys, so that we simply state that the results are in good
agreement.

A more quantitative comparison is possible by comparing our survey to the
photometric COMBO-17 survey \citep{wolf03}. Since COMBO-17 is flux-limited in
the $R$ band whereas our survey is best defined in $V$, we assumed a mean quasar
colour of $V-R=0.2$ to convert the COMBO-17 surface densities. After correction
for the selection function, the cumulative surface number density of quasars
found for our survey is 48 deg$^{-2}$ at $V<21$, and 89 deg$^{-2}$ at $V<22$.
The corresponding values for COMBO-17 are 51.3 deg$^{-2}$ and 93.5 deg$^{-2}$,
respectively. Thus again, the numbers are in excellent agreement, even at
magnitudes as faint as $V\sim 22$. Looking closer at the \emph{differential}
number counts we find that they agree very well at $20.5 < V < 21.5$
($42 \pm 8$ in our survey vs.\ $46\pm 8$ in COMBO-17), but that our survey drops
by a factor of 1.5 in the subsequent magnitude bin at $21.5 < V < 22.5$
($41 \pm 11$ vs.\ $63\pm 9$). It is not surprising to find that our survey is
incomplete at these magnitudes, probably as a combination of incompleteness
already in the input catalogues and missing true quasars among the noisy spectra.

As we had included the ECDFS from COMBO-17 in our survey observations, we could
also perform an object by object comparison in this well-studied field. The
slitless spectra confirmed the general trend of the above survey selection
function. We rediscovered 7 out of 10 $R<21$ COMBO-17 quasars at $1.3<z<3.6$
located in the common survey area. The three missed quasars are quite faint
($R>20$) and reside at $z\la 1.9$, where the selection function of our survey
drops (Fig.~\ref{qnq_completeness4}). At $z>2.5$ only one out of three $21<R<22$
quasars was rediscovered, indicating photometric incompleteness at the faint
end of the DSS catalogue.

\subsection{Comparison to other surveys for quasar groups}

Slitless spectroscopic surveys are successful in finding closely projected
quasar groups at high redshift, especially in redshift ranges where colour
surveys are less effective. However, previous targeted slitless surveys for
quasar groups were few, always pointed at one individual field only
\citep[e.g.][]{jakobsen92,williger96}, and also generally shallower than our
survey by $\sim 1$~mag. Thus, the resulting surface maps of quasars were sparse,
but probed the large-scale distribution of quasars. Our survey significantly
increases the number of high-redshift quasar groups with separations
$\vartheta\la 15\arcmin$ in the southern hemisphere. The centroiding of our
fields on known bright quasars enables follow-up studies on the
three-dimensional distribution of the IGM. Our faint survey limit
($V_\mathrm{lim}=$21--22) results in a high surface density of high-redshift
quasars despite the comparatively small survey area per field.

More recently, \citet{hennawi06b} selected faint $i\la 21$ quasar candidates in
the vicinity of confirmed SDSS quasars using the SDSS photometry. Their survey
was dedicated to find close quasar pairs with separations
$\vartheta\la 1\arcmin$, especially binary quasars at the same redshift. They
confirmed 40 new binary quasars and 73 projected quasar pairs. However, 89 of
their candidates turned out to be stars, most likely due to broadband colour
similarities between stars and quasars in certain redshift ranges. In
comparison, our high success rate of almost 100\% shows that slitless
spectroscopy is highly efficient in finding high-redshift quasars. A smooth
selection function (Fig.~\ref{qnq_completeness4}) is essential to discover
high-redshift quasar groups in selected fields. However, given the small
surface density of high-redshift quasars, we were unlikely to find quasars at
very small separations to the central quasars in our fields.

\section{Conclusions}
\label{qnq_conclusions}

We performed the ``Quasars near Quasars'' (QNQ) survey, a CCD-based slitless
spectroscopic survey for faint $V\la 22$ quasars at $1.7\la z\la 3.6$ around
18 well-studied bright quasars at $2.76<z<4.69$, covering a total area of
$\simeq 4.39$~deg$^2$. In order to analyse the slitless data we developed a data
reduction pipeline that performs an optimal extraction of the spectra and that
is able to cope with contaminating spectral orders of slitless grism data. From
the $\sim 29000$ extracted flux-calibrated spectra we selected 169 previously
unknown quasar candidates and $>100$ likely low-redshift emission line galaxy
candidates on the basis of emission features that fall in the covered wavelength
range 4200~\AA$\le\lambda\le$5800~\AA. A semi-automatic selection routine
limited potential biases of purely visual selection and allowed to quantify
selection effects.

Follow-up spectroscopy confirmed 80 out of 81 selected quasar candidates on 16
fields. 64 of these newly established quasars reside at $z>1.7$. The highest
redshift quasar is QNQ~J22484$-$6002 at $z=3.586$. The brightest newly
discovered high-redshift quasar is QNQ~J11197$-$1340 ($z=2.220$, $B=18.3$).
Given the high success rate of the follow-up, the vast majority of the remaining
88 candidates will be quasars as well, although most of them likely reside at
lower redshifts. 

The primary aim of this survey was to provide new groups of quasars for
medium-resolution spectroscopy in the vicinity of well-studied known quasars.
Originally, we had not expected to reach as faint as $V\sim 22$ in the survey
observations. At these faint magnitudes our survey is not well defined because
of photometric incompleteness in the source catalogues needed for automatic
extraction of the slitless spectra. Together with the fact that not all selected
quasar candidates could be included in the spectroscopic follow-up, our survey
probably is of limited use for constraining quasar evolution. However, focusing
on survey efficiency rather than on completeness at the faint end was justified
in order to accomplish the main goal of this survey. In fact, the faintest
quasars discovered will for now remain beyond the limits for obtaining
high-quality spectra. But at least the $V\la 21$ quasars are well suited for
follow-up studies with current (e.g.\ ESI at Keck) or upcoming
(e.g.\ X-shooter at VLT) high-throughput spectrographs at 8--10~m-class telescopes.

Together with the central quasars in the fields already observed at high
resolution, these quasar groups can be used to perform a tomography of the
intergalactic medium. Large-scale clustering of the Ly$\alpha$ forest or
correlations of metal line systems can be investigated as well
\citep{williger00,dodorico02,dodorico06}. Some of the discovered quasars reside
at similar redshifts or approximately at the same redshift of the central quasar
in the field, giving potentially insights to quasar clustering along overdense
filaments in the plane of the sky. We have identified two quasars coinciding
with a damped Ly$\alpha$ absorber on one central line of sight, as well as a
large-scale group of quasars at $z=2.70$--$2.78$. Moreover, our study provides
new foreground quasars to investigate the transverse proximity effect of quasars
(Papers~I and II). In combination with the already available high-resolution
spectra of the central quasars, medium-resolution spectra of these quasar groups
will offer great opportunities to study the large-scale cosmic web in three
dimensions.

\begin{acknowledgements}
We thank the staff of the ESO observatories La Silla and Paranal for their
professional assistance in obtaining the data presented in this paper.
This research has made use of the NASA/IPAC Extragalactic Database (NED) which
is operated by the Jet Propulsion Laboratory, California Institute of Technology,
under contract with the National Aeronautics and Space Administration. G.W. was
partly supported by a HWP grant from the state of Brandenburg, Germany. G.W. and
L.W. acknowledge support by the Deutsche Forschungsgemeinschaft under Wi~1369/21-1.
\end{acknowledgements}

\bibliography{qnqpaper}

\begin{thebibliography}{94}
\expandafter\ifx\csname natexlab\endcsname\relax\def\natexlab#1{#1}\fi

\bibitem[{Adelberger {et~al.}(2005)Adelberger, Shapley, Steidel,
  {et~al.}}]{adelberger05}
Adelberger, K.~L., Shapley, A.~E., Steidel, C.~C., {et~al.} 2005, ApJ, 629, 636

\bibitem[{Appenzeller {et~al.}(1998)Appenzeller, Fricke, Furtig,
  {et~al.}}]{appenzeller98}
Appenzeller, I., Fricke, K., Furtig, W., {et~al.} 1998, The Messenger, 94, 1

\bibitem[{Baade {et~al.}(1999)Baade, Meisenheimer, Iwert, {et~al.}}]{baade99}
Baade, D., Meisenheimer, K., Iwert, O., {et~al.} 1999, The Messenger, 95, 15

\bibitem[{Barbieri \& Cristiani(1986)}]{barbieri86}
Barbieri, C. \& Cristiani, S. 1986, A\&AS, 63, 1

\bibitem[{Barger {et~al.}(2003)Barger, Cowie, Capak, {et~al.}}]{barger03}
Barger, A.~J., Cowie, L.~L., Capak, P., {et~al.} 2003, AJ, 126, 632

\bibitem[{Becker {et~al.}(2007)Becker, Rauch, \& Sargent}]{becker07}
Becker, G.~D., Rauch, M., \& Sargent, W.~L.~W. 2007, ApJ, 662, 72

\bibitem[{Bergeron {et~al.}(2004)Bergeron, Petitjean, Aracil,
  {et~al.}}]{bergeron04}
Bergeron, J., Petitjean, P., Aracil, B., {et~al.} 2004, The Messenger, 118, 40

\bibitem[{Bohuski \& Weedman(1979)}]{bohuski79}
Bohuski, T.~J. \& Weedman, D.~W. 1979, ApJ, 231, 653

\bibitem[{Bongiovanni {et~al.}(2005)Bongiovanni, Bruzual, Magris,
  {et~al.}}]{bongiovanni05}
Bongiovanni, A., Bruzual, G., Magris, G., {et~al.} 2005, MNRAS, 359, 930

\bibitem[{Chand {et~al.}(2004)Chand, Srianand, Petitjean, \& Aracil}]{chand04}
Chand, H., Srianand, R., Petitjean, P., \& Aracil, B. 2004, A\&A, 417, 853

\bibitem[{Ciliegi {et~al.}(1994)Ciliegi, Elvis, Gioia, Maccacaro, \&
  Wolter}]{ciliegi94}
Ciliegi, P., Elvis, M., Gioia, I.~M., Maccacaro, T., \& Wolter, A. 1994, AJ,
  108, 970

\bibitem[{Clowes {et~al.}(1999)Clowes, Campusano, \& Graham}]{clowes99}
Clowes, R.~G., Campusano, L.~E., \& Graham, M.~J. 1999, MNRAS, 309, 48

\bibitem[{Clowes {et~al.}(1984)Clowes, Cooke, \& Beard}]{clowes84}
Clowes, R.~G., Cooke, J.~A., \& Beard, S.~M. 1984, MNRAS, 207, 99

\bibitem[{Cowie {et~al.}(2004)Cowie, Barger, Hu, Capak, \& Songaila}]{cowie04}
Cowie, L.~L., Barger, A.~J., Hu, E.~M., Capak, P., \& Songaila, A. 2004, AJ,
  127, 3137

\bibitem[{Crampton {et~al.}(1990)Crampton, Cowley, \& Hartwick}]{crampton90}
Crampton, D., Cowley, A.~P., \& Hartwick, F.~D.~A. 1990, AJ, 100, 47

\bibitem[{Crampton {et~al.}(1985)Crampton, Schade, \& Cowley}]{crampton85}
Crampton, D., Schade, D., \& Cowley, A.~P. 1985, AJ, 90, 987

\bibitem[{Croom {et~al.}(2004)Croom, Smith, Boyle, {et~al.}}]{croom04}
Croom, S.~M., Smith, R.~J., Boyle, B.~J., {et~al.} 2004, MNRAS, 349, 1397

\bibitem[{Crotts \& Fang(1998)}]{crotts98}
Crotts, A.~P.~S. \& Fang, Y. 1998, ApJ, 502, 16

\bibitem[{Dinshaw \& Impey(1996)}]{dinshaw96}
Dinshaw, N. \& Impey, C.~D. 1996, ApJ, 458, 73

\bibitem[{{D'Odorico} {et~al.}(2002){D'Odorico}, Petitjean, Cristiani, \&
  {D'Odorico}}]{dodorico02}
{D'Odorico}, V., Petitjean, P., Cristiani, S., \& {D'Odorico}, S. 2002, A\&A,
  390, 13

\bibitem[{{D'Odorico} {et~al.}(2006){D'Odorico}, Viel, Saitta,
  {et~al.}}]{dodorico06}
{D'Odorico}, V., Viel, M., Saitta, F., {et~al.} 2006, MNRAS, 372, 1333

\bibitem[{Francis {et~al.}(2004)Francis, Palunas, Teplitz, Williger, \&
  Woodgate}]{francis04}
Francis, P.~J., Palunas, P., Teplitz, H.~I., Williger, G.~M., \& Woodgate,
  B.~E. 2004, ApJ, 614, 75

\bibitem[{Gaskell(1982)}]{gaskell82}
Gaskell, C.~M. 1982, ApJ, 263, 79

\bibitem[{Gratton \& Osmer(1987)}]{gratton87}
Gratton, R.~G. \& Osmer, P.~S. 1987, PASP, 99, 899

\bibitem[{Hawkins(2000)}]{hawkins00}
Hawkins, M.~R.~S. 2000, A\&AS, 143, 465

\bibitem[{Hazard \& McMahon(1985)}]{hazard85}
Hazard, C. \& McMahon, R. 1985, Nat, 314, 238

\bibitem[{Hazard {et~al.}(1987)Hazard, McMahon, \& Morton}]{hazard87}
Hazard, C., McMahon, R.~G., \& Morton, D.~C. 1987, MNRAS, 229, 371

\bibitem[{Hazard {et~al.}(1986)Hazard, Morton, McMahon, Sargent, \&
  Terlevich}]{hazard86}
Hazard, C., Morton, D.~C., McMahon, R.~G., Sargent, W.~L.~W., \& Terlevich, R.
  1986, MNRAS, 223, 87

\bibitem[{Heap {et~al.}(2000)Heap, Williger, Smette, {et~al.}}]{heap00}
Heap, S.~R., Williger, G.~M., Smette, A., {et~al.} 2000, ApJ, 534, 69

\bibitem[{Hennawi \& Prochaska(2007)}]{hennawi07}
Hennawi, J.~F. \& Prochaska, J.~X. 2007, ApJ, 655, 735

\bibitem[{Hennawi {et~al.}(2006{\natexlab{a}})Hennawi, Prochaska, Burles,
  {et~al.}}]{hennawi06a}
Hennawi, J.~F., Prochaska, J.~X., Burles, S., {et~al.} 2006{\natexlab{a}}, ApJ,
  651, 61

\bibitem[{Hennawi {et~al.}(2006{\natexlab{b}})Hennawi, Strauss, Oguri,
  {et~al.}}]{hennawi06b}
Hennawi, J.~F., Strauss, M.~A., Oguri, M., {et~al.} 2006{\natexlab{b}}, AJ,
  131, 1

\bibitem[{Hewett {et~al.}(1995)Hewett, Foltz, \& Chaffee}]{hewett95}
Hewett, P.~C., Foltz, C.~B., \& Chaffee, F.~H. 1995, AJ, 109, 1498

\bibitem[{Hewett {et~al.}(1985)Hewett, Irwin, Bunclark, {et~al.}}]{hewett85}
Hewett, P.~C., Irwin, M.~J., Bunclark, P., {et~al.} 1985, MNRAS, 213, 971

\bibitem[{Hewitt \& Burbidge(1993)}]{hewitt93}
Hewitt, A. \& Burbidge, G. 1993, ApJS, 87, 451

\bibitem[{Hook {et~al.}(2003)Hook, Shaver, Jackson, Wall, \&
  Kellermann}]{hook03}
Hook, I.~M., Shaver, P.~A., Jackson, C.~A., Wall, J.~V., \& Kellermann, K.~I.
  2003, A\&A, 399, 469

\bibitem[{Horne(1986)}]{horne86}
Horne, K. 1986, PASP, 98, 609

\bibitem[{Jakobsen {et~al.}(1994)Jakobsen, Boksenberg, Deharveng,
  {et~al.}}]{jakobsen94}
Jakobsen, P., Boksenberg, A., Deharveng, J.~M., {et~al.} 1994, Nat, 370, 35

\bibitem[{Jakobsen {et~al.}(2003)Jakobsen, Jansen, Wagner, \&
  Reimers}]{jakobsen03}
Jakobsen, P., Jansen, R.~A., Wagner, S., \& Reimers, D. 2003, A\&A, 397, 891

\bibitem[{Jakobsen \& Perryman(1992)}]{jakobsen92}
Jakobsen, P. \& Perryman, M.~A.~C. 1992, ApJ, 392, 432

\bibitem[{Jakobsen {et~al.}(1986)Jakobsen, Perryman, Ulrich, Macchetto, \& {di
  Serego Alighieri}}]{jakobsen86}
Jakobsen, P., Perryman, M.~A.~C., Ulrich, M.~H., Macchetto, F., \& {di Serego
  Alighieri}, S. 1986, ApJ, 303, L27

\bibitem[{Jauncey {et~al.}(1978)Jauncey, Wright, Peterson, \&
  Condon}]{jauncey78}
Jauncey, D.~L., Wright, A.~E., Peterson, B.~A., \& Condon, J.~J. 1978, ApJ,
  219, L1

\bibitem[{Jiang {et~al.}(2008)Jiang, Fan, Annis, {et~al.}}]{jiang08}
Jiang, L., Fan, X., Annis, J., {et~al.} 2008, AJ, 135, 1057

\bibitem[{Kim {et~al.}(2007)Kim, Bolton, Viel, Haehnelt, \& Carswell}]{kim07b}
Kim, T.-S., Bolton, J.~S., Viel, M., Haehnelt, M.~G., \& Carswell, R.~F. 2007,
  MNRAS, 382, 1657

\bibitem[{Krumpe {et~al.}(2007)Krumpe, Lamer, Schwope, {et~al.}}]{krumpe07}
Krumpe, M., Lamer, G., Schwope, A.~D., {et~al.} 2007, A\&A, 466, 41

\bibitem[{{La Franca} {et~al.}(1999){La Franca}, Lissandrini, Cristiani,
  {et~al.}}]{lafranca99}
{La Franca}, F., Lissandrini, C., Cristiani, S., {et~al.} 1999, A\&AS, 140, 351

\bibitem[{Landolt(1992)}]{landolt92}
Landolt, A.~U. 1992, AJ, 104, 340

\bibitem[{Liske {et~al.}(2000)Liske, Webb, Williger, Fern\'{a}ndez-Soto, \&
  Carswell}]{liske00b}
Liske, J., Webb, J.~K., Williger, G.~M., Fern\'{a}ndez-Soto, A., \& Carswell,
  R.~F. 2000, MNRAS, 311, 657

\bibitem[{Lopez {et~al.}(2002)Lopez, Reimers, {D'Odorico}, \&
  Prochaska}]{lopez02}
Lopez, S., Reimers, D., {D'Odorico}, S., \& Prochaska, J.~X. 2002, A\&A, 385,
  778

\bibitem[{Maza {et~al.}(1993)Maza, Ruiz, Gonz\'{a}lez, Wischnjewsky, \&
  Antezana}]{maza93}
Maza, J., Ruiz, M.~T., Gonz\'{a}lez, L.~E., Wischnjewsky, M., \& Antezana, R.
  1993, RMxAA, 25, 51

\bibitem[{Maza {et~al.}(1995)Maza, Wischnjewsky, Antezana, \&
  Gonz\'{a}lez}]{maza95}
Maza, J., Wischnjewsky, M., Antezana, R., \& Gonz\'{a}lez, L.~E. 1995, RMxAA,
  31, 119

\bibitem[{McIntosh {et~al.}(1999)McIntosh, Rix, Rieke, \& Foltz}]{mcintosh99}
McIntosh, D.~H., Rix, H.-W., Rieke, M.~J., \& Foltz, C.~B. 1999, ApJ, 517, L73

\bibitem[{McMahon {et~al.}(1994)McMahon, Omont, Bergeron, Kreysa, \&
  Haslam}]{mcmahon94}
McMahon, R.~G., Omont, A., Bergeron, J., Kreysa, E., \& Haslam, C.~G.~T. 1994,
  MNRAS, 267, L9

\bibitem[{Morris {et~al.}(1991)Morris, Weymann, Anderson, {et~al.}}]{morris91}
Morris, S.~L., Weymann, R.~J., Anderson, S.~F., {et~al.} 1991, AJ, 102, 1627

\bibitem[{Murphy {et~al.}(2003)Murphy, Webb, \& Flambaum}]{murphy03}
Murphy, M.~T., Webb, J.~K., \& Flambaum, V.~V. 2003, MNRAS, 345, 609

\bibitem[{Osmer(1980)}]{osmer80a}
Osmer, P.~S. 1980, ApJS, 42, 523

\bibitem[{Osmer \& Smith(1976)}]{osmer76}
Osmer, P.~S. \& Smith, M.~G. 1976, ApJ, 210, 267

\bibitem[{Osmer \& Smith(1980)}]{osmer80b}
Osmer, P.~S. \& Smith, M.~G. 1980, ApJS, 42, 333

\bibitem[{Pascual {et~al.}(2001)Pascual, Gallego, Arag{\'o}n-Salamanca, \&
  Zamorano}]{pascual01}
Pascual, S., Gallego, J., Arag{\'o}n-Salamanca, A., \& Zamorano, J. 2001, A\&A,
  379, 798

\bibitem[{Pichon {et~al.}(2001)Pichon, Vergely, Rollinde, Colombi, \&
  Petitjean}]{pichon01}
Pichon, C., Vergely, J.~L., Rollinde, E., Colombi, S., \& Petitjean, P. 2001,
  MNRAS, 326, 597

\bibitem[{Prescott {et~al.}(2006)Prescott, Impey, Cool, \&
  Scoville}]{prescott06}
Prescott, M.~K.~M., Impey, C.~D., Cool, R.~J., \& Scoville, N.~Z. 2006, ApJ,
  644, 100

\bibitem[{Reimers {et~al.}(1997)Reimers, K\"{o}hler, Wisotzki,
  {et~al.}}]{reimers97}
Reimers, D., K\"{o}hler, S., Wisotzki, L., {et~al.} 1997, A\&A, 327, 890

\bibitem[{Reimers {et~al.}(1995)Reimers, Rodriguez-Pascual, Hagen, \&
  Wisotzki}]{reimers95}
Reimers, D., Rodriguez-Pascual, P., Hagen, H.-J., \& Wisotzki, L. 1995, A\&A,
  293, L21

\bibitem[{Richards {et~al.}(2002{\natexlab{a}})Richards, Fan, Newberg,
  {et~al.}}]{richards02b}
Richards, G.~T., Fan, X., Newberg, H.~J., {et~al.} 2002{\natexlab{a}}, AJ, 123,
  2945

\bibitem[{Richards {et~al.}(2002{\natexlab{b}})Richards, {Vanden Berk},
  Reichard, {et~al.}}]{richards02}
Richards, G.~T., {Vanden Berk}, D.~E., Reichard, T.~A., {et~al.}
  2002{\natexlab{b}}, AJ, 124, 1

\bibitem[{Salzer {et~al.}(2002)Salzer, Gronwall, Sarajedini,
  {et~al.}}]{salzer02}
Salzer, J.~J., Gronwall, C., Sarajedini, V.~L., {et~al.} 2002, AJ, 123, 1292

\bibitem[{Schmidt {et~al.}(1986)Schmidt, Schneider, \& Gunn}]{schmidt86a}
Schmidt, M., Schneider, D.~P., \& Gunn, J.~E. 1986, ApJ, 306, 411

\bibitem[{Schmidt {et~al.}(1995)Schmidt, Schneider, \& Gunn}]{schmidt95}
Schmidt, M., Schneider, D.~P., \& Gunn, J.~E. 1995, AJ, 110, 68

\bibitem[{Schneider {et~al.}(2007)Schneider, Hall, Richards,
  {et~al.}}]{schneider07}
Schneider, D.~P., Hall, P.~B., Richards, G.~T., {et~al.} 2007, AJ, 134, 102

\bibitem[{Schneider {et~al.}(1994)Schneider, Schmidt, \& Gunn}]{schneider94}
Schneider, D.~P., Schmidt, M., \& Gunn, J.~E. 1994, AJ, 107, 1245

\bibitem[{Schneider {et~al.}(1999)Schneider, Schmidt, \& Gunn}]{schneider99}
Schneider, D.~P., Schmidt, M., \& Gunn, J.~E. 1999, AJ, 117, 40

\bibitem[{Sharp {et~al.}(2002)Sharp, Sabbey, Vivas, {et~al.}}]{sharp02}
Sharp, R.~G., Sabbey, C.~N., Vivas, A.~K., {et~al.} 2002, MNRAS, 337, 1153

\bibitem[{Shaver(1987)}]{shaver87}
Shaver, P. 1987, Nat, 330, 426

\bibitem[{Smette {et~al.}(2002)Smette, Heap, Williger, {et~al.}}]{smette02}
Smette, A., Heap, S.~R., Williger, G.~M., {et~al.} 2002, ApJ, 564, 542

\bibitem[{Sramek \& Weedman(1978)}]{sramek78}
Sramek, R.~A. \& Weedman, D.~W. 1978, ApJ, 221, 468

\bibitem[{Steidel {et~al.}(2003)Steidel, Adelberger, Shapley,
  {et~al.}}]{steidel03}
Steidel, C.~C., Adelberger, K.~L., Shapley, A.~E., {et~al.} 2003, ApJ, 592, 728

\bibitem[{Storrie-Lombardi {et~al.}(1996)Storrie-Lombardi, McMahon, Irwin, \&
  Hazard}]{storrie-lombardi96}
Storrie-Lombardi, L.~J., McMahon, R.~G., Irwin, M.~J., \& Hazard, C. 1996, ApJ,
  468, 121

\bibitem[{Szokoly {et~al.}(2004)Szokoly, Bergeron, Hasinger,
  {et~al.}}]{szokoly04}
Szokoly, G.~P., Bergeron, J., Hasinger, G., {et~al.} 2004, ApJS, 155, 271

\bibitem[{Trump {et~al.}(2007)Trump, Impey, McCarthy, {et~al.}}]{trump07}
Trump, J.~R., Impey, C.~D., McCarthy, P.~J., {et~al.} 2007, ApJS, 172, 383

\bibitem[{Tytler \& Fan(1992)}]{tytler92}
Tytler, D. \& Fan, X. 1992, ApJS, 79, 1

\bibitem[{{Vanden Berk} {et~al.}(2001){Vanden Berk}, Richards, Bauer,
  {et~al.}}]{vandenberk01}
{Vanden Berk}, D.~E., Richards, G.~T., Bauer, A., {et~al.} 2001, AJ, 122, 549

\bibitem[{V\'{e}ron \& Hawkins(1995)}]{veron95}
V\'{e}ron, P. \& Hawkins, M.~R.~S. 1995, A\&A, 296, 665

\bibitem[{Warren {et~al.}(1991)Warren, Hewett, Irwin, \& Osmer}]{warren91a}
Warren, S.~J., Hewett, P.~C., Irwin, M.~J., \& Osmer, P.~S. 1991, ApJS, 76, 1

\bibitem[{Williger {et~al.}(1996)Williger, Hazard, Baldwin, \&
  McMahon}]{williger96}
Williger, G.~M., Hazard, C., Baldwin, J.~A., \& McMahon, R.~G. 1996, ApJS, 104,
  145

\bibitem[{Williger {et~al.}(2000)Williger, Smette, Hazard, Baldwin, \&
  McMahon}]{williger00}
Williger, G.~M., Smette, A., Hazard, C., Baldwin, J.~A., \& McMahon, R.~G.
  2000, ApJ, 532, 77

\bibitem[{Wisotzki {et~al.}(2000)Wisotzki, Christlieb, Bade,
  {et~al.}}]{wisotzki00}
Wisotzki, L., Christlieb, N., Bade, N., {et~al.} 2000, A\&A, 358, 77

\bibitem[{Wisotzki {et~al.}(2001)Wisotzki, Selman, \& Gilliotte}]{wisotzki01}
Wisotzki, L., Selman, F., \& Gilliotte, A. 2001, The Messenger, 104, 8

\bibitem[{Wolf {et~al.}(2004)Wolf, Meisenheimer, Kleinheinrich,
  {et~al.}}]{wolf04}
Wolf, C., Meisenheimer, K., Kleinheinrich, M., {et~al.} 2004, A\&A, 421, 913

\bibitem[{Wolf {et~al.}(2003)Wolf, Wisotzki, Borch, {et~al.}}]{wolf03}
Wolf, C., Wisotzki, L., Borch, A., {et~al.} 2003, A\&A, 408, 499

\bibitem[{Worseck {et~al.}(2007)Worseck, Fechner, Wisotzki, \&
  {Dall'~Aglio}}]{worseck07}
Worseck, G., Fechner, C., Wisotzki, L., \& {Dall'~Aglio}, A. 2007, A\&A, 473,
  805

\bibitem[{Worseck \& Wisotzki(2006)}]{worseck06}
Worseck, G. \& Wisotzki, L. 2006, A\&A, 450, 495

\bibitem[{Wright {et~al.}(1977)Wright, Jauncey, Peterson, \& Condon}]{wright77}
Wright, A.~E., Jauncey, D.~L., Peterson, B.~A., \& Condon, J.~J. 1977, ApJ,
  211, L115

\bibitem[{Zheng {et~al.}(2004)Zheng, Kriss, Deharveng, {et~al.}}]{zheng04}
Zheng, W., Kriss, G.~A., Deharveng, J.-M., {et~al.} 2004, ApJ, 605, 631

\bibitem[{Zitelli {et~al.}(1992)Zitelli, Mignoli, Zamorani, Marano, \&
  Boyle}]{zitelli92}
Zitelli, V., Mignoli, M., Zamorani, G., Marano, B., \& Boyle, B.~J. 1992,
  MNRAS, 256, 349

\end{thebibliography}

\end{document}